\documentclass[twocolumn, showpacs,superscriptaddress, nofootinbib]{revtex4-2}
\usepackage{multirow}
\usepackage{mathrsfs}
\usepackage{amsmath}
\usepackage[colorlinks=true, allcolors=blue]{hyperref}
\usepackage{verbatim}
\usepackage{amsfonts}
\usepackage{amssymb}
\usepackage{amsthm}
\usepackage{bm}
\usepackage{dsfont}
\usepackage{xparse}
\usepackage{graphicx}
\usepackage{xcolor}
\usepackage{textcase}
\usepackage{url}
\usepackage{lipsum}
\usepackage{epstopdf}
\usepackage{footnote}
\usepackage{color}  

\newtheorem*{lemma*}{Lemma}

\def\a{\alpha}
\def\b{\beta}

\def\G{\Gamma}

\def\e{\epsilon}
\def\l{\lambda}

\def\r{\rho}
\def\s{\sigma}

\def\j{\varphi}
\def\f{\phi}

\def\tr{\text{tr}}

\def\ox{\otimes}

\def\var{\text{Var}}
\def\cov{\text{Cov}}
\def\re{\text{Re}}
\def\im{\text{Im}}

\newcommand{\avg}[1]{\left\langle{#1}\right\rangle}
\newcommand{\ket}[1]{| {#1} \rangle} 
\newcommand{\bra}[1]{\langle {#1} |} 

\newcommand{\mel}[3]{\langle {#1} \vphantom{#2} | {#2} \vphantom{#3} | {#3} \rangle}
 
\DeclareMathOperator*{\Var}{\mathrm{Var}}
\DeclareMathOperator*{\E}{\mathbb{E}}
\DeclareMathOperator*{\Cov}{\mathrm{Cov}}

\usepackage{enumitem}
\usepackage{float}

\begin{document}
\title{Robust and Indestructible Macroscopic Entanglement}
\author{Martina Gisti}
\email{martina.gisti@uni-bonn.de}
\affiliation{University of Vienna, Faculty of Physics, Vienna Center for Quantum Science and Technology, Boltzmanngasse 5, 1090 Vienna, Austria}
\affiliation{Institute of Physics, University of Bonn, Nußallee 12, 53115 Bonn, Germany}
\author{Miguel Gallego}
\email{miguel.gallego.ballester@univie.ac.at}
\affiliation{University of Vienna, Faculty of Physics, Vienna Center for Quantum Science and Technology, Boltzmanngasse 5, 1090 Vienna, Austria}
\affiliation{University of Vienna, Vienna Doctoral School in Physics, Bolztmanngasse 5, 1090 Vienna, Austria}
\author{Borivoje Daki\'c}
\email{borivoje.dakic@univie.ac.at}
\affiliation{University of Vienna, Faculty of Physics, Vienna Center for Quantum Science and Technology, Boltzmanngasse 5, 1090 Vienna, Austria}
\affiliation{Institute for Quantum Optics and Quantum Information (IQOQI), Austrian Academy of Sciences, Boltzmanngasse 3, A-1090 Vienna, Austria}

\date{\today}

\begin{abstract}
In this letter we investigate the possibility of observing macroscopic entanglement, considering realistic factors such as decoherence, particle losses, and measurements of limited precision (coarse-grained collective measurements). This forms the foundation for an operational definition of robust macroscopic entanglement. We also propose an even more resilient concept of macroscopic entanglement, allowing the parties to randomly select their portion of the shared system. Through a straightforward example, we demonstrate that achieving this is feasible using a simple class of quantum states. Our result opens a path for generalizing notions of quantum entanglement, such as a genuine multipartite entanglement in the macroscopic domain.
\end{abstract}

\maketitle

\section*{Introduction}
Dating back to Schrödinger's famous cat paradox \cite{schrodinger}, macroscopic quantum entanglement is a fascinating phenomenon that challenges the applicability of quantum mechanics at large scales, where classical behavior dominates. In the past decades, we have seen a significant technological advancement to enable the manipulation and control of quantum systems with a large number of degrees of freedom (typically particle number), with notable examples including spin-squeezed atomic ensembles and Bose-Einstein condensate (BEC) systems \cite{Julsgaard_2001,Leroux2010,hosten2016measurement,gross2011atomic}, or systems of mechanical oscillators \cite{thomas, mercier, kotler, direkci}. Despite these achievements, pushing the limits to even larger scales is challenging due to the quantum-to-classical transition, a boundary where quantum systems effectively start behaving classically. From a dynamical point of view, this transition occurs due to decoherence \cite{zurek1, zurek2}, based on the experimental difficulty of isolating a macroscopic system from its environment. However, even in perfect isolation conditions, limited measurement precision washes out all quantumness in the system due to the coarse-graining mechanism ~\cite{koflerbrukner,kofler2013}. While coarse-graining is typically attributed to the `kinematical side' of the transition to classicality, it can be directly related to decoherence (dynamics) through the modeling of the measurement process where both can be seen as an instance of the same phenomenon \cite{gallego}. In any case, these effects put limitations and practical restrictions on detecting macroscopic entanglement. To illustrate this point, we refer to a typical situation where quantum entanglement of a macroscopic system is inferred indirectly via the witness method, where a suitably chosen set of collective (additive) observables is measured to detect entanglement \cite{guhne2008,Toth05}. For example, consider the measurement of total angular momentum for spin systems $A^{(N)} = \frac{1}{N^\alpha}\sum_{k=1}^N A_k$, with $A_k$ being a single-particle binary observable (spin along some direction, with two possible values $\pm1$). In such a case, the observable $A^{(N)}$ takes values in the interval $[-N/N^\a,+N/N^\a]$. Thus, the size of the measurement slot (difference between neighboring outcomes) amounts to $1/N^\alpha$, which is referred to as the measurement precision. The constant $\alpha\in[0,1]$ is the coarse-graining factor \cite{gallegodakic}. Now, for two collective observables $A^{(N)}$ and $B^{(N)}$ their commutator evaluates to 
\begin{equation}
    [A^{(N)},B^{(N)}]=i C^{(N)}=\frac{i}{N^{2\alpha}}\sum_{k=1}^N C_k,
\end{equation}
with $C_k=-i[A_k,B_k]$. We see that the operator norm of the commutator $||C^{(N)}||=O(1/N^{2\alpha-1})$, thus $A^{(N)}$ and $B^{(N)}$ become commutative in the macroscopic limit ($N \to \infty$) for $\alpha>1/2$. In such a case, all collective observables become effectively compatible in the limit and entanglement cannot be detected based on their measurement. Similar reasoning can be found in the context of uncertainty relations \cite{demelo}. Although the presented argument is not rigorous, it illustrates well the findings of \cite{koflerbrukner,kofler2013}, where a measurement precision much smaller than $1/\sqrt N$ effectively leads to statistics which can be reproduced by classical systems. An interesting case is $\alpha=1/2$, which has been shown to exhibit local correlations (in the sense of Bell's theorem) for a simple class of macroscopic states, namely states consisting of many \emph{independent and identically distributed} (IID) correlated pairs. This has been know in the literature as the principle of \emph{macroscopic locality}~\cite{navascues2010glance}. However, if we abandon the IID assumption, the system can display macroscopically nonlocal quantum correlations~\cite{gallegodakic, gallego, gallegodakic2}. Beyond the works concentrating on device-independent tests of quantumness in the macroscopic system \cite{jeong2014, gallegodakic}, a considerable body of work has been done in a sole quantum setting with the focus on standard entanglement witness methods~\cite{guhne2008}. Nevertheless, in most of the studies, the focus is on entanglement detection between microscopic constituents via collective observables measured on a whole system \cite{bruknervedral, bruknervedralzeilinger, wiesniakmagnetic, wiesniakheat,rosario,rosario19, rosario20, rosario21, rosario22,rosario23,rosario24,rosario25,rosario26,rosario27}.
In contrast, in this work, we focus on entanglement detection where the macroscopic systems are divided into two (macroscopic) parts on which the collective measurements are locally performed. This makes a more clear operational route towards defining the notion of a \emph{robust macroscopic entanglement}. We then go further and introduce the even stronger notion of \emph{indestructible macroscopic entanglement}, which is a type of robust macroscopic entanglement that can be witnessed even when the parties do not have control over the partition of the macroscopic system and thus select their parts at random. We provide examples of simple and experimentally friendly witnesses that can be used to detect such entanglement by reformulating the celebrated Duan-Simon's criterion \cite{duan, simon} to our scenario. Surprisingly, such entanglement can be found in simple classes of quantum states such as the IID entangled pairs. These results shed new light on the question of the quantum-to-classical transition, suggesting that entanglement might be much more robust than expected (as compared, for example, to Bell's nonlocality).

\section*{Robust Macroscopic Entanglement (RME)}
\label{MQE}
\begin{figure}
        \centering
        \includegraphics[clip, trim=0.5cm 3.5cm 0.5cm 8cm, width=0.35\textwidth]{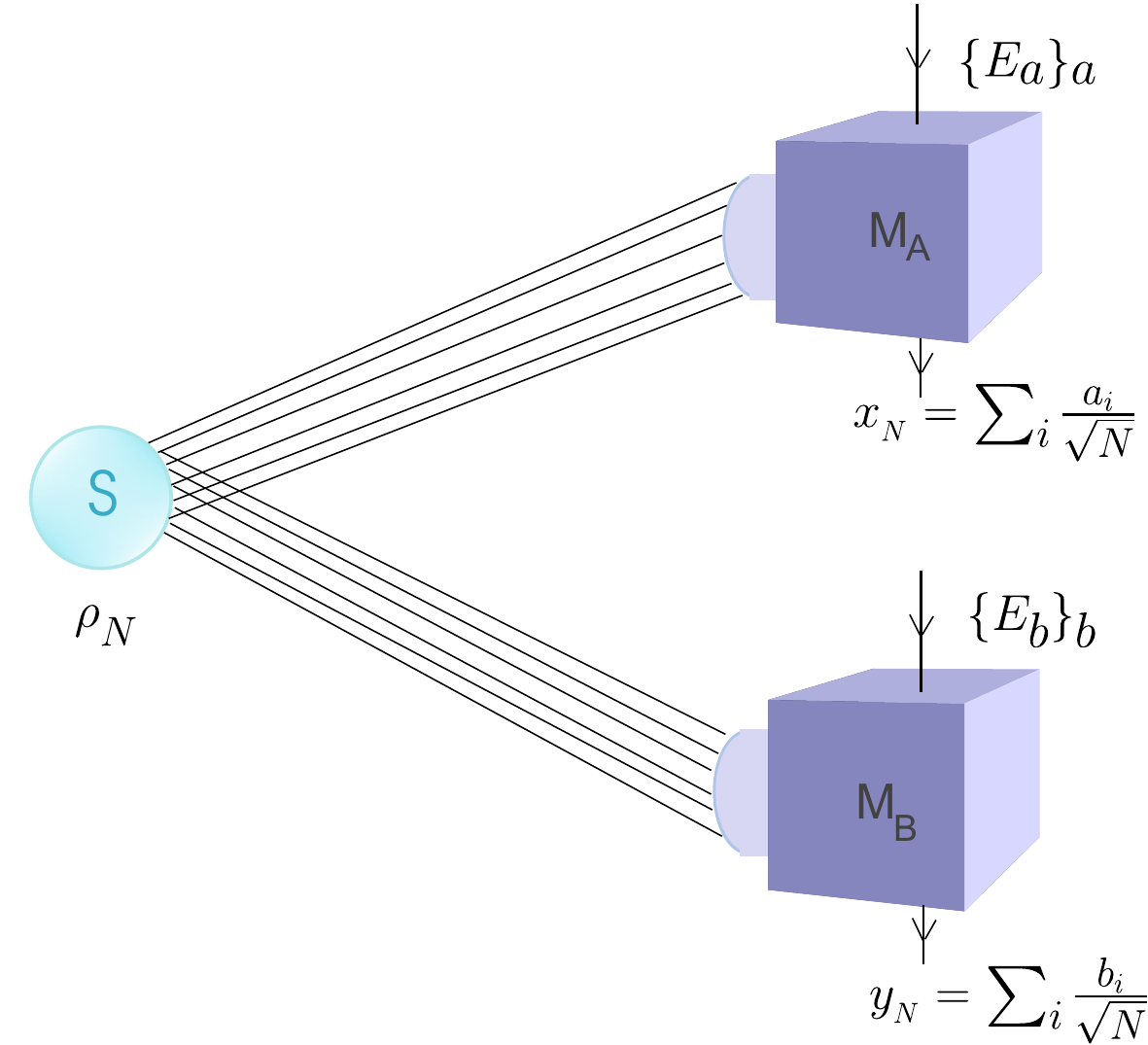}
        \caption{\textbf{Macroscopic measurement.} An $N$-particle quantum system $\mathsf{S}$ described by a state $\r_N$ subject to the assumptions \ref{ass:largen} - \ref{ass:loss} is divided in two parts $\mathsf{S}_A$ and $\mathsf{S}_B$ which are sent to the measurement apparatuses $\mathsf{M}_A$ and $\mathsf{M}_B$ respectively, both subject to assumptions \ref{ass:coll} - \ref{ass:cg}.}
        \label{fig:meas}
\end{figure}

We now present an operational definition of macroscopic entanglement based on the observability of entanglement in the macroscopic limit under the effects of decoherence, particle losses, and coarse-grained measurements. For this purpose, we consider a macroscopic quantum measurement scenario similar to the one in references \cite{nw, gallegodakic, gallego, gallegodakic2}. For simplicity, we restrict to the case of bipartite entanglement, the generalization to the multipartite case being straightforward (see figure \ref{fig:meas}). A quantum system $\mathsf{S}$ consisting of a large number $N$ of particles or subsystems is prepared in some quantum state. The system is subject to the effects of independent single-particle decoherence (through some single-particle decoherence channel, such as the depolarizing channel) and particle losses (i.e., each particle has some probability of being lost before it can be measured). The system is then divided into two parts $\mathsf{S}_A$ and $\mathsf{S}_B$, which are sent to respective measurement apparatuses $\mathsf{M}_A$ and $\mathsf{M}_B$. Each measurement apparatus implements a collective, coarse-grained intensity measurement. Namely, apparatus $\mathsf{M}_A$ measures a quantity of the form $\sum_i a_i$ (where $a_i$ is the outcome of a measurement on the $i$-th particle of $\mathsf{S}_A$) with a \emph{precision} or \emph{resolution} of the order of $\sqrt{N}$. Similarly, apparatus $\mathsf{M}_B$ measures $\sum_i b_i$ (where $b_i$ is the outcome of a measurement on the $i$-th particle of $\mathsf{S}_B$) with resolution of order $\sqrt{N}$. We will then say that the state of the system possesses \emph{robust macroscopic entanglement} (RME) if its entanglement is \emph{observable} or \emph{detectable} in the above circumstances in the limit of infinitely many particles. In the next section, we mathematically formalize this definition. It is worth pointing out that we have set explicitly the coarse-graining factor $\alpha=1/2$ (measurement precision of the order $1/\sqrt N$), for which our main results are derived. Nevertheless, in the outlook section we comment on the possible extensions to arbitrary $\alpha\in[0,1]$.

\section*{Mathematical Framework for RME}
Let us make precise the assumptions involved in the macroscopic quantum measurement scenario described in the previous section, i.e., let us assume that the quantum system $\mathsf{S}$ satisfies the following conditions:
\begin{enumerate}[label=(\textit{\roman*})]
    \item \emph{Large $N$.} The system is composed of a large number $N$ of identical particles or subsystems.\footnote{The assumption that the subsystems be identical is made for the sake of simplicity} We associate a Hilbert space $\mathcal{H}$ to each particle, so that the Hilbert space of the whole system is $\mathcal{H}^{\ox N}$, which is bipartitioned into Alice's Hilbert space $\mathcal{H}_A = \mathcal{H}^{\ox N_A}$ and Bob's $\mathcal{H}_B = \mathcal{H}^{\ox N_B}$, where $N_A + N_B = N$ (i.e., $\mathcal{H}^{\ox N} \cong \mathcal{H}_A \ox \mathcal{H}_B$). The state of the system is described by a density matrix $\rho_N \in \mathcal{D}(\mathcal{H}^{\otimes N})$, i.e. a positive, self-adjoint bounded linear operator on $\mathcal{H}^{\otimes N}$ satisfying $\tr \, \rho_N =1$.
    \label{ass:largen}

    \item \emph{Decoherence.} The system is subject to independent, single-particle decoherence channels $\Gamma$. The effective state is therefore $\Gamma^{\otimes N} ( \rho_N )$.
    \label{ass:decoh}

    \item \emph{Particle losses.} Each individual particle is lost with probability $p \in [0,1]$, while $1-p$ is its probability of reaching the measurement apparatuses. Therefore, on each run of the experiment only a number $M \leq N$ of particles actually reaches the measurement apparatuses, and this occurs with probability $P_N(M) = {N \choose M} p^{N-M} (1-p)^{M}$, in which case the effective state of the system is of the form  $\tr_{N-M} (\r_N)$.
    \label{ass:loss}
\end{enumerate}
Putting together these assumptions, we have that for a given initial $N$-particle state $\r_N \in \mathcal{D}(\mathcal{H}^{\ox N})$, the effective state of the system can be written in the `Fock space' $\oplus_{M=0}^N \mathcal{H}^{\ox M}$ as
\begin{align}
    \r_N^{\text{eff}} = \bigoplus_{M=0}^N  P_N(M) \,  \sum_{\pi \in \mathfrak{S}_N} \frac{\tr_{\pi(1) \dots \pi(N-M)} \Big[  \G^{\otimes N} ( \rho_N )  \Big] }{N!}  \, ,
\label{eq:rhoneff}
\end{align}
where $\mathfrak{S}_N$ is the symmetric group. 


As for the measurement, we assume that Alice's measurement apparatus $\mathsf{M}_A$ satisfies the following conditions:
\begin{enumerate}[label=(\textit{\roman*})]
    \setcounter{enumi}{3}
    \item \emph{Collective measurement}. The measurement apparatus implements the same \emph{positive operator-valued measure} (POVM) on each particle or subsystem. To each outcome $a \in \mathbb{R}$ of the POVM is associated a positive operator $E_a  \in \mathcal{B}(\mathcal{H})$. This leads to the following joint distribution of outcomes
    \begin{align}
        P(a_1, \dots, a_N) = \tr \Big[ \r_N  E_{a_1} \ox \dots \ox E_{a_N} \Big] \, .
    \end{align}
    \label{ass:coll}
    
    \item \emph{Intensity measurement.} The measurement apparatus has no access to individual single-particle outcomes $a$, but only to their sum $\sum_{i} a_i$ (we leave the limits of the sum unspecified because the number of particles in the effective state $\r_N^{\text{eff}}$ is indeterminate). 
    \label{ass:intensity}
    
    \item \emph{Coarse-graining.} The measuring scale for the intensity $\sum_{i} a_i$ has a limited resolution of the order of $\sqrt{N}$, meaning that it cannot distinguish between values that differ by much less than $\sqrt{N}$. The accessible quantity is therefore $x_N = \sum_i a_i/\sqrt{N}$. For a discussion of how the coarse-graining assumption in the limit $N \to \infty$ is implemented by dividing by a factor of $\sqrt{N}$, see \cite{gallego}.
    \label{ass:cg}
\end{enumerate}
The same limitations apply to Bob's measurement apparatus $\mathsf{M}_B$, whose accessible quantity is thus $y_N = \sum_i b_i /\sqrt{N}$. 

As outlined before, we say that a state $\r_N$ possesses RME if its entanglement is observable in the above scenario in the limit $N \to \infty$, with the state being subject to assumptions \ref{ass:largen} - \ref{ass:loss} and the measurements being subject to assumptions \ref{ass:coll} - \ref{ass:cg}. More explicitly, $\r_N$ possesses RME if there exists a set of measurements and experimental procedures consistent with \ref{ass:largen} - \ref{ass:cg} such that we can detect entanglement, i.e. such that from the collected statistics we can show that
\begin{align*}
\r_N\neq\sum_s\lambda_s\r_{N_A}^{(s)}\otimes\r_{N_B}^{(s)} \, .
\end{align*}
In order for the robustness conditions to be satisfied, this means that there exists $\epsilon > 0$ and $\delta > 0$ such that for all $\G$ satisfying $\| \G - \text{Id} \| < \epsilon$ and for all $p < \delta$, the entanglement of $\r_N^{\text{eff}}$ is observable with measurements of coarse-grained intensities $x_N$ and $y_N$. It is worth pointing out that these are necessary for robustness. Of course, as we shall see, RME can exist even in more `noisy scenarios,' such as $p$ being significantly larger than 0. To illustrate this idea, let us examine some examples and non-examples. First, consider a bipartite state where all but a pair of particles are in a separable state, for instance, a state of the form $\ket{\psi}_{AB} \ox (\ket{\f}_A \ox \ket{\chi}_B )^{\ox (N-1)}$, where $\ket{\psi}_{AB}$ is the state of a pair of particles shared by the parties and $\ket{\f}_A$ and $\ket{\chi}_B$ are the states of the rest of Alice's and Bob's particles respectively. Then, only the first pair of particles is entangled, and its contribution to the quantities $x_N$ and $y_N$ vanishes in the limit $N \to  \infty$, so that no entanglement can be observed in the limit. This is a case of microscopic entanglement between macroscopic systems that fails to satisfy our definition of RME. Next, consider a state chosen at random according to the Haar measure in $\mathcal{H}^{\ox N}$. It has been shown \cite{bremner, gross} that this kind of state possesses a large amount of entanglement, but complicated measurements are required to observe it. We therefore expect the requirements of RME to fail. Another non-example is the Greenberger-Horne-Zeilinger \cite{ghz} state $( \ket{0}^{\ox N} + \ket{1}^{\ox N} ) / \sqrt{2}$. This state is typically associated with macroscopic entanglement when $N$ is large. Still, it is extremely fragile since the loss of coherence of a single particle destroys the coherence of the state altogether, collapsing it into a classical mixture. Indeed, we can see that such a state does not possess RME as we have defined it since all the components of its effective state \eqref{eq:rhoneff} are separable except for the $N$-particle component, but this component has a weight $P_N(N) = (1-p)^N$ which is vanishingly small in the macroscopic limit for all $p>0$. The other state typically associated with macroscopic entanglement is the W \cite{w} state $(|10 \dots 0\rangle +|01 \dots 0\rangle + \dots +|00 \dots 1\rangle ) / {\sqrt{N}}$. We do expect this state to constitute an example of RME. However, we claim that a simpler state is sufficient to exhibit RME, namely an IID state, i.e. a state of the form $\ket{\psi}^{\ox N}$, where $\ket{\psi}$ is an entangled two-particle state shared by Alice and Bob. We now prove RME for such a state.

\subsection*{An example of RME}
\label{RME exampe}

For this purpose, let us first introduce a slight generalization of the Duan-Simon entanglement criterion \cite{duan, simon}. Consider a bipartite quantum state $\r \in \mathcal{H}_A \ox \mathcal{H}_B$, where $\mathcal{H}_A$ and $\mathcal{H}_B$ are arbitrary Hilbert spaces, and suppose Alice and Bob both measure one of two observables, say $x_A$ and $p_A$ for Alice and $x_B$ and $p_B$ for Bob (these are arbitrary observables, i.e. not necessarily position and momentum observables as in the Duan-Simon criterion). Then, as we show in Appendix \ref{app:duan}, all separable states $\r$ satisfy $f \geq 0$ with
\begin{align}
    f & = \var (x_A + x_B) + \var (p_A - p_B)  \nonumber \\
    & \qquad \qquad \quad  - \big| \langle [x_A, p_A] \rangle \big| - \big| \langle [x_B, p_B] \rangle \big| \, , 
\label{eq:duansimon}
\end{align}
where $\langle \xi \rangle = \tr ( \r \, \xi )$ and $\var(\xi) = \langle \xi^2 \rangle - \langle \xi \rangle^2$. Now, consider the case where $\mathcal{H}_A = \mathcal{H}_B = \mathcal{H}^{\ox N}$ for some Hilbert space $\mathcal{H}$, so that $\r$ can be seen as a $2N$-particle state $\r_{2N} \in \mathcal{D} (\mathcal{H}^{\ox N} \ox \mathcal{H}^{\ox N})$. Furthermore, suppose that the observables measured by the parties are coarse-grained intensities as in assumptions \ref{ass:coll} - \ref{ass:cg}, i.e. suppose that
\begin{align*}
x_A = \sum_{i=1}^{N} \frac{A_1^{(i)} }{\sqrt{N}} \, , \qquad p_A = \sum_{i=1}^{N} \frac{A_2^{(i)} }{\sqrt{N}} \, ,  \\
x_B = \sum_{i=1}^{N} \frac{B_1^{(i)} }{\sqrt{N}} \, , \qquad p_B = \sum_{i=1}^{N} \frac{B_2^{(i)}}{\sqrt{N}} \, ,
\end{align*}
for some single-particle observables $A_1$ and $A_2$ for Alice and $B_1$ and $B_2$ for Bob. In this case, a violation of the inequality $f \geq 0$ would be a sufficient criterion for RME. If the state is IID of the form $\r_{2N} = (\ket{\psi} \bra{\psi} )^{\ox N}$, where $\ket{\psi}$ is the state of two entangled particles shared by the two parties, then, as we show in Appendix \ref{app:duanIID}, the function $f$ can be written without dependence on $N$ as
\begin{align}
    f & = \var(A_1+B_1) + \var(A_2-B_2)  \nonumber \\
    & \qquad \qquad \quad - \big| \langle [A_1, A_2] \rangle \big| - \big| \langle [B_1, B_2] \rangle \big| \, ,
\label{eq:duansimoniid}
\end{align}
where now $\langle X \rangle = \mel{\psi}{X}{\psi}$ and $\var(X) = \langle X^2 \rangle - \avg{X}^2$. In order to compute the maximal violation of the inequality, notice that its potential violation is unbounded since the function $f$ is homogeneous in the observables $A_1$, $A_2$, $B_1$, and $B_2$. It therefore makes sense to restrict the norm of these to be bounded, so we set $\|A_1\|, \|A_2\|, \|B_1\|, \|B_2\|\leq1$. Then, for the simple case where $\dim \mathcal{H} = 2$, we numerically compute the maximal violation over states $\ket{\psi}$ and spin observables $A_1$, $A_2$, $B_1$ and $B_2$, and recognize the analytical value
\begin{align*}
    f = 4(1 - \sqrt{2}) \simeq -1.66
\end{align*}
for
\begin{align}
   \ket{\psi} = \cos(\pi/8) \ket{00} - \sin(\pi/8) \ket{11}
   \label{eq:IIDstateRME}
\end{align}
and $A_1= B_1= \s_x$ and $A_2=B_2=\s_y$, where $\s_k$ is the $k$-th Pauli matrix. Note that the state $\ket{\psi}$ that gives the maximal violation for the entanglement inequality is not maximally entangled and, as a matter of fact, we did not find any violation of the inequality for maximally entangled two-qubit states. 

In Appendix \ref{app:RME} we quantify the robustness of the violation against decoherence, particle losses and non-projective measurements. Fixing the state and the observables that give the maximal violation in the perfect case, the violation is still visible if the state is affected by the IID decoherence channel $\G^{\ox N}$ with $\G(\ket{\psi} \bra{\psi}) = (1-\l)\ket{\psi} \bra{\psi} + \l \, \mathbb{I}_2 /2$ being the depolarizing channel with 
\begin{align*}
    \l < \frac{3-\sqrt{1+4\sqrt{2}}}{2} \simeq 0.21 \, .
\end{align*}
If the state is affected by particle losses, with probability $p$ of being lost, the violation is visible as long as 
\begin{align*}
    p < 2 - \sqrt{2} \simeq 0.59 \, .
\end{align*}
And if instead of the observables $A_1$, $A_2$, $B_1$ and $B_2$ some other POVMs $\epsilon$ away are actually performed, the violation is still visible if
\begin{align*}
    \epsilon < \frac{2-\sqrt{2}}{3 \sqrt{2}} \simeq 0.10 \, .
\end{align*}

\section*{Indestructible Macroscopic Entanglement (IME)}
Let us now introduce a stronger notion of macroscopic entanglement than RME, where the parties need not have perfect control (actually need not have control at all) over the exact bipartition of the Hilbert space $\mathcal{H}^{\ox 2N} \cong \mathcal{H}^{\ox N} \ox \mathcal{H}^{\ox N}$, which in the previous section was assumed to be given. We will say that a state $\r_{2N}$ possesses \emph{indestructible macroscopic entanglement} (IME) if its entanglement is observable in the limit $N \to \infty$ in the defined macroscopic scenario (i.e., subject to assumptions \ref{ass:largen} - \ref{ass:cg}) \emph{for a (average) random bipartition of the Hilbert space $\mathcal{H}^{\ox 2N}$}. This definition resembles genuine multipartite entanglement~\cite{guhne2008} but adapted to the macroscopic limit. While it may seem quite demanding and one might think it is theoretically empty, we surprisingly demonstrate that simple IID states are sufficient to meet the conditions for IME. Let us prove this.

\begin{figure}
    \centering
    \includegraphics[width=0.45\textwidth]{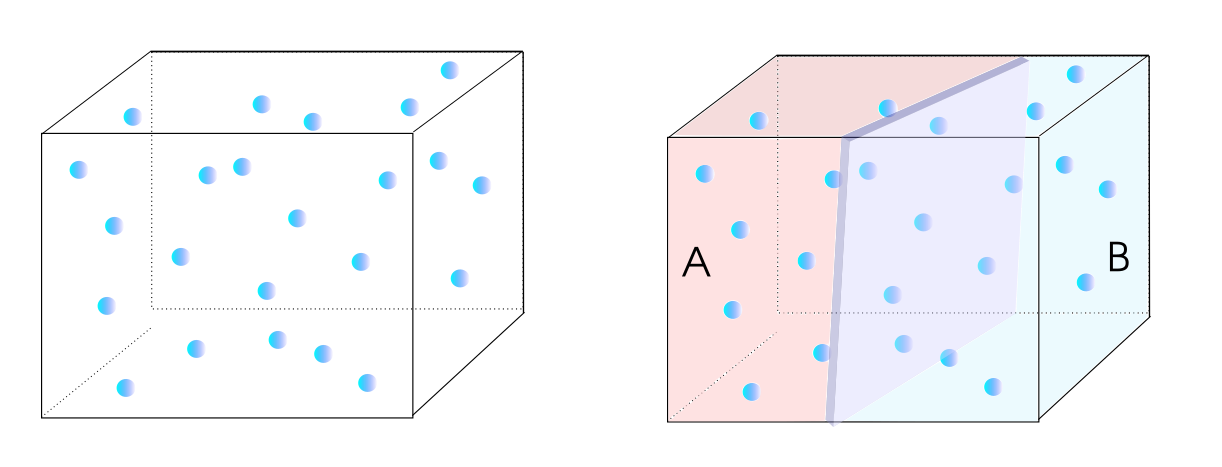}
    \caption{\textbf{Random bipartition.} A random bipartition $\mathcal{H}^{\ox 2N} \cong \mathcal{H}_A \otimes \mathcal{H}_B $ is modeled by distributing the position of each particle uniformly inside the volume of a box and dividing in two parts corresponding to fractions $q$ and $(1-q)$ of the total volume, where $q$ is chosen uniformly at random in the unit interval. }
\label{fig:randombipartition}
\end{figure}

\begin{figure}
    \centering
    \includegraphics[width=1.\columnwidth]{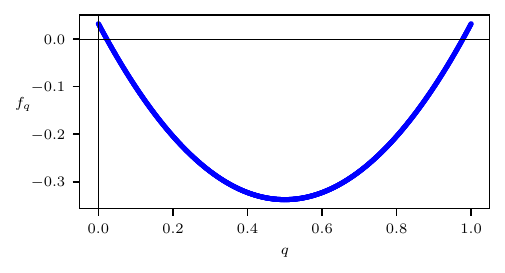}
    \caption{The inequality $f_q \geq 0$ is violated for most $q$ in $[0,1]$, leading to the violation of the inequality on average.}
\label{fig:plot}
\end{figure}

For that purpose, we will proceed as in the previous section, showing a violation of the inequality $f\geq 0$ with $f$ as in \eqref{eq:duansimon}, but now using the state resulting from a random bipartition of the IID state $\ket{\psi}^{\ox N}$ with $\ket{\psi} \in \mathcal{H} \ox \mathcal{H}$. To model a random bipartition, we assume each particle has a probability $q \in [0,1]$ of falling on one side, say Alice's, while $1-q$ is the probability that it falls on Bob's side (we can call this a \emph{$q$-bipartition}), and then we average over $q$. One can picture this procedure by imagining the $2N$ particles being uniformly distributed (in position) inside a box that we divide in two parts with volume fractions $q$ and $1-q$, where $q$ is chosen uniformly at random from the unit interval (see figure \ref{fig:randombipartition}). Let us now first focus on a $q$-bipartition with $q$ fixed. For simplicity, let us assume that the state $\ket{\psi}$ is permutationally invariant, so that we do not have to know which particle of the pair is which. Then, for a given $q$, the probability that a number $N_A \leq N$ of entangled pairs falls on Alice's side and that a number $N_B \leq N - N_A$ falls on Bob's side is: 
\begin{widetext}
\begin{align*}
    P_{N}(q,N_A,N_B) = \frac{N!}{N_A! N_{AB}! N_B!} \, q^{ 2 N_A} \big( 2 q (1-q) \big)^{N_{AB}} (1-q)^{2 N_B} ,
\end{align*}
\end{widetext}
where $N_{AB} = N - N_A - N_B$ is the number of entangled pairs that are divided across the bipartition (no matter how). In such a configuration, the state is
\begin{align*}
\r_{2N}(N_A,N_B) = \s_{A A}^{\otimes N_A} \ox \s_{A B}^{\otimes N_{AB}} \ox \s_{B B}^{\otimes N_B} \, ,
\end{align*}
where $\s = \ket{\psi} \bra{\psi}$ and $\s_{AA}$ represents the state of a pair of particles in the state $\s$ falling on Alice's side, $\s_{BB}$ the state of a pair falling on Bob's side and $\s_{AB}$ the state of a pair shared between the two parties. The effective state of the system is thus
\begin{align}
\r_{2N}(q) = \sum_{N_A = 0}^{N}  \sum_{N_B = 0}^{N-N_A} P_{N}(q,N_A,N_B) \cdot \r_{2N}(N_A,N_B) \, .
\label{eq:rho2nq}
\end{align}
If Alice measures one of two observables, either $x_A$ or $p_A$, and Bob measures $x_B$ or $p_B$, with
\begin{align*}
x_A = \sum_{i=1}^{2 N_A +N_{AB}} \frac{A_1^{(i)} }{\sqrt{N}} \, , \qquad p_A = \sum_{i=1}^{2 N_A + N_{AB}} \frac{A_2^{(i)} }{\sqrt{N}} \, ,\\
x_B = \sum_{i=1}^{2 N_B +N_{AB}} \frac{B_1^{(i)} }{\sqrt{N}} \, , \qquad p_B = \sum_{i=1}^{2 N_B +N_{AB}} \frac{B_2^{(i)}}{\sqrt{N}} \, ,
\end{align*}
as we show in Appendix \ref{app:duanrandombipartition}, the function $f$ can be written as
\begin{align}
    f_q = &  2 q \Big(  \avg{A_1^2}  + q \avg{A_1 \ox A_1}  - 2q \avg{A_1}^2 \Big)          \nonumber  \\
    & +  2\bar{q} \Big( \avg{B_1^2} + \bar{q} \avg{B_1 \ox B_1} - 2\bar{q} \avg{B_1}^2 \Big)   \nonumber  \\
    & + 4q \bar{q} \Big( \avg{A_1 \ox B_1} - 2 \avg{A_1} \avg{ B_1} \Big)      \nonumber  \\
    & +  2q \Big( \avg{A_2^2}  + q \avg{A_2 \ox A_2} - 2q \avg{A_2}^2 \Big)          \nonumber  \\
    & + 2 \bar{q} \Big( \avg{B_2^2}  + \bar{q} \avg{B_2 \ox B_2} - 2\bar{q} \avg{B_2}^2 \Big) \nonumber  \\
    & -  4 q \bar{q} \Big( \avg{A_2 \ox B_2} - 2 \avg{A_2} \avg{B_2} \Big)     \nonumber  \\ 
    & - 2 q \, \big| \langle [A_1 , A_2] \rangle \big| -  2\bar{q} \big| \langle [B_1 , B_2] \rangle  \big|  \, .
\label{eq:fq}
\end{align}
where $\langle X \rangle=\tr ( \s X )$ and $\bar{q} = 1-q$. Now, for the inequality $f_q \geq 0$ we found no qubit violation (i.e. no violation for $\dim \mathcal{H} = 2$). However, if $\dim \mathcal{H}=3$,  an appropriate choice of the state $\ket{\psi} \in \mathbb{C}^3$ and of the observables $\{ A_1, B_1, A_2, B_2\} \subseteq \mathcal{B}(\mathbb{C}^3)$ gives numerical violation of the inequality $f_q \geq 0$ for almost all $q \in [0,1]$ (see Figure \ref{fig:plot}) . Moreover, a single choice of state and observables gives a violation of the inequality \emph{even for average} $q$, which in turn means that the state $(\ket{\psi} \bra{\psi})^{\ox N}$ possesses IME as we have defined it. To see this, consider the average over $q$ of Equation \eqref{eq:fq}:
\begin{align}
    f = &    \avg{A_1^2} + \frac{2}{3} \avg{A_1 \ox A_1} - \frac{4}{3} \avg{A_1}^2     \nonumber  \\
    & +    \avg{B_1^2} + \frac{2}{3} \avg{B_1 \ox B_1}   - \frac{4}{3} \avg{B_1}^2\nonumber  \\
    & + \frac{2}{3} \Big( \avg{A_1 \ox B_1} - 2 \avg{A_1} \avg{ B_1} \Big)      \nonumber  \\
    & +  \avg{A_2^2} + \frac{2}{3} \avg{A_2 \ox A_2} - \frac{4}{3} \avg{A_2}^2      \nonumber  \\
    & +   \avg{B_2^2}  + \frac{2}{3} \avg{B_2 \ox B_2}- \frac{4}{3} \avg{B_2}^2 \nonumber  \\
    & - \frac{2}{3} \Big( \avg{A_2 \ox B_2} - 2 \avg{A_2} \avg{B_2} \Big)     \nonumber  \\ 
    & -  \, \big| \langle [A_1 , A_2] \rangle \big| -  \big| \langle [B_1 , B_2] \rangle  \big|  \, .
\label{eq:favgq}
\end{align}
Then (restricting the norm of the single-particle observables to be less or equal than $1$), the maximal violation we found for states $\s = \ket{\psi} \bra{\psi}$ and for spin observables $A_1$, $A_2$, $B_1$ and $B_2$ is
\begin{align*}
    f \simeq -0.21
\end{align*}
for $\psi = \sum_{i,j=0}^2 c_{ij} \ket{ij}$ with
\begin{align}
    c_{00} & \simeq 0.34 - 0.87 i \, , \nonumber \\
    c_{02} & \simeq 0.07 \, , \nonumber \\
    c_{11} & \simeq - 0.33 \, , \nonumber \\
    c_{22} & \simeq 0.03 + 0.07 i \, ,
\label{eq:cijnumerical}
\end{align}
(all other coefficients being zero) and for $A_j = \cos \j_j^{(A)} S_x + \sin \j_j^{(A)} S_y$ and $B_j = \cos \j_j^{(B)} S_x + \sin \j_j^{(B)} S_y$ with
\begin{align}
    \j_1^{(A)} & = 0 \, ,  \nonumber \\
    \j_2^{(A)} & =: \f \simeq 1.20  \, , \nonumber \\
    \j_1^{(B)} & = \f \simeq 1.20 \, , \nonumber \\
    \j_2^{(B)}  & = \pi \, ,
\label{eq:anglesnumerical}
\end{align}
where
\begin{align*}
    S_x = \frac{1}{\sqrt{2}} \begin{pmatrix}
        0 & 1 & 0 \\
        1 & 0 & 1 \\
        0 & 1 & 0
    \end{pmatrix} \, , \quad 
    S_y = \frac{1}{\sqrt{2}}
    \begin{pmatrix}
        0 & -i & 0 \\
        i & 0 & -i \\
        0 & i & 0
    \end{pmatrix} \, .
\end{align*}

In Appendix \ref{app:IMErobustness} we quantify the robustness of the violation (fixing the state and observables) against the effects of decoherence, losses and non-projective measurements. First, consider the effect of depolarization, captured by the single-particle channel $\G(\s) = (1-\lambda) \s + \lambda \, \mathbb{I}_{3}/3$ for $\lambda \in [0,1]$. Then the inequality $f\geq 0$ is violated as long as
\begin{align*}
    \l \lesssim 0.06 \, . 
\end{align*}
Next, in order to consider the effect of particle losses, we assume each particle is lost with probability $p \in [0,1]$, reaching the detector with probability $1-p$. Then, the averaged inequality is violated for
\begin{align*}
    p \lesssim 0.24 .
\end{align*}
Finally, consider the case where the measurement apparatus do not implement single-particle projective measurements but POVMs. Namely, let Alice measure $x_A = \sum_{i} {a_i }/{\sqrt{N}}$
where each $a$ is the outcome of a single-particle POVM with elements $E_a$, and suppose $||E_a - \Pi_a || < \epsilon $ (similarly for Bob). Then, the inequality is violated for 
\begin{align*}
    \epsilon \lesssim 0.02 .
\end{align*}
This shows the robustness of the macroscopic entanglement for the state considered, thus proving IME.

\section*{Outlook}
We have proposed an operational definition of macroscopic entanglement under the realistic assumptions \ref{ass:largen} - \ref{ass:cg}. We have introduced two notions of macroscopic entanglement: RME and IME. The latter assembles ideas of genuine multiparty entanglement~\cite{guhne2008} adapted to such macroscopic measurements and robust scenarios. An interesting feature is that a non-linear witness \eqref{eq:duansimoniid} was necessary to find the entanglement criterion, as we failed to find an example with a linear witness. A similar result has been observed for continuous-variable Bell's inequalities \cite{hecavalcantireiddrummond}, and this finding is definitely worth investigating further. Another relevant point to be investigated is the level of coarse-graining (or measurement precision), which we have set to the order of $\sqrt{N}$ (i.e. $\alpha=1/2$). The witness equation \eqref{eq:duansimoniid} is homogenous, so it can be easily extended to arbitrary $\alpha$, yielding
\begin{align*}
    f & = \frac{1}{N^{2 \a - 1}} \Big[ \var(A_1+B_1) + \var(A_2-B_2)  \nonumber \\
    & \qquad \qquad \quad - \big| \langle [A_1, A_2] \rangle \big| - \big| \langle [B_1, B_2] \rangle \big| \Big] \, .
\end{align*}
What we see is that $f$ diverges (as $N \to \infty$) for $\alpha<1/2$, while for $\alpha>1/2$ we have $f\to 0$, meaning that entanglement cannot be detected in the macroscopic limit via the witness \eqref{eq:duansimoniid} in such case. This coincides with the results of \cite{koflerbrukner,kofler2013,gallegodakic,gallegodakic2} for device-independent quantum correlations, which transit to classical ones in this regime. While this is true for quantum nonlocality as a resource, it does not have to be necessarily true for entanglement. Thus, the question remains open whether quantum entanglement can still be detected in the macroscopic regime for $\alpha>1/2$.\\

\emph{Acknowledgments.}--- This research was funded in whole, or in part, by the Austrian Science Fund (FWF) [10.55776/F71] and [10.55776/P36994].  Martina Gisti acknowledges support of the Erasmus Programme.  Miguel Gallego acknowledges support from the ESQ Discovery programme (Erwin Schrödinger Center for Quantum Science \& Technology), hosted by the Austrian Academy of Sciences (OAW). For open access purposes, the author(s) has applied a CC BY public copyright license to any author accepted manuscript version arising from this submission.

\bibliographystyle{apsrev4-1}
\bibliography{References}{}

\onecolumngrid
\appendix

\section{Generalization of the Duan-Simon entanglement criterion}
\label{app:duan}
Consider the bipartite Hilbert space $\mathcal{H}_A \ox \mathcal{H}_B$ and let $\r \in \mathcal{D}(\mathcal{H}_A \ox \mathcal{H}_B)$ be a separable state, i.e.
 \begin{align}
 \rho = \sum_i p_i \, \rho_A^{(i)} \otimes \rho_B^{(i)} 
\end{align}   
for some $p_i \geq 0$ satisfying $\sum_i p_i = 1$ and for some density matrices $\{ \r_A^{(i)} \}_i \subset \mathcal{D}(\mathcal{H}_A)$ and $\{ \r_B^{(i)} \}_i \subset \mathcal{D}(\mathcal{H}_B)$ . Let $x_A$ and $p_A$ be two arbitrary observables on $\mathcal{H}_A$, and let $x_B$ and $p_B$ be two arbitrary observables on $\mathcal{H}_B$. Then, letting $\langle x \rangle = \tr ( \r x )$ and $\langle x \rangle_i = \tr ( \r_A^{(i)} \ox \r_A^{(i)} x ) $ we have
\begin{align}
\Var (x_A +x_B)   & = \avg{ (x_A +x_B)^2}-  \avg{x_A + x_B}^2 \nonumber \\
& = \sum_i p_i \avg{ (x_A +x_B)^2}_i -  \avg{x_A + x_B}^2 \nonumber \\
& = \sum_i p_i \Big( \avg{ x_A^2}_i + \avg{ x_B^2}_i + 2 \, \avg{ x_A  x_B}_i  \Big)  -  \avg{ x_A +x_B}^2 \nonumber \\
& = \sum_i p_i \Big( \avg{ x_A^2}_i + \avg{ x_B^2}_i + 2 \, \avg{ x_A}_i \avg{x_B}_i  \Big)  -  \avg{ x_A +x_B}^2 \nonumber \\
& = \sum_i p_i \Big( \avg{ x_A^2}_i - \avg{x_A}_i^2 + \avg{ x_B^2}_i - \avg{x_B}_i^2 \Big) + \sum_i p_i \Big( \avg{x_A}_i^2 + \avg{x_B}_i^2 + 2 \, \avg{ x_A}_i \avg{x_B}_i \Big)  -  \avg{ x_A +x_B}^2 \nonumber \\
& = \sum_i p_i \Big( {\Var}_i ( x_A)+ {\Var}_i (x_B)\Big) + \sum_i p_i \big( \avg{x_A}_i + \avg{x_B}_i \big)^2  -  \avg{ x_A +x_B}^2 \nonumber \\
& \geq \sum_i p_i \Big( {\Var}_i ( x_A)+ {\Var}_i (x_B)\Big) \, ,
\end{align}
where $\Var_i (x) = \tr ( \r_A^{(i)} \ox \r_A^{(i)} x^2 ) - \tr ( \r_A^{(i)} \ox \r_A^{(i)} x )^2$ and in the last line we have used that $\sum_i p_i {\langle x \rangle_i}^2 \geq \left( \sum_i p_i | \langle x \rangle_i | \right)^2 \geq \langle x \rangle^2$. Similarly, 
\begin{align}
    \Var (p_A -p_B) \geq   \sum_i p_i \Big( {\Var}_i ( p_A)+ {\Var}_i (p_B)\Big) \, .
\end{align}
Hence
\begin{align}
    \Var (x_A +x_B) + \Var (p_A -p_B) & \geq \sum_i p_i \Big( {\Var}_i ( x_A)+ {\Var}_i (x_B) +  {\Var}_i ( p_A)+ {\Var}_i (p_B) \Big) \nonumber \\
    &  = \sum_i p_i \Bigg[  \left( \sqrt{{\Var}_i ( x_A)} - \sqrt{{\Var}_i ( p_A)} \right)^2 +2\sqrt{{\Var}_i ( x_A)} \sqrt{{\Var}_i ( p_A)} \nonumber \\
    & \qquad \qquad +  \left( \sqrt{{\Var}_i ( x_B)} - \sqrt{{\Var}_i ( p_B)} \right)^2 +2\sqrt{{\Var}_i ( x_B)} \sqrt{{\Var}_i ( p_B)}  \Bigg] \nonumber \\
    & \geq \sum_i p_i\left( 2 \sqrt{{\Var}_i ( x_A)} \sqrt{{\Var}_i ( p_A)}  +  2\sqrt{{\Var}_i ( x_B)} \sqrt{{\Var}_i ( p_B)} \right) \nonumber \\
    & \geq \sum_i p_i\left(\big| \langle [x_A, p_A] \rangle_i \big| + \big| \langle [x_B, p_B] \rangle_i \big|   \right) \nonumber \\
    & \geq \big| \langle [x_A, p_A] \rangle \big| + \big| \langle [x_B, p_B] \rangle \big| \, ,
\label{eq:app:duansimon}
\end{align}
where in the second to last line we have used the uncertainty principle \cite{robertson}, proving Equation \eqref{eq:duansimon} in the main text.

\section{Inequality for IID states and coarse-grained collective observables}
\label{app:duanIID}
We can rewrite the inequality \eqref{eq:app:duansimon} as
\begin{align}
    \Var(x_A) + \Var(x_B) + 2 \Cov (x_A, x_B) +  \Var(p_A) + \Var(p_B) - 2 \Cov (p_A, p_B) \geq  \big| \langle [x_A, p_A] \rangle \big| + \big| \langle [x_B, p_B] \rangle \big| \, ,
\end{align}
where $\Cov(x,y) = \langle x \ox y \rangle - \langle x \rangle \langle  y \rangle$. If the state is IID, i.e. $\r_{2N} = ( | \psi \rangle \langle \psi | ) ^{\ox N}$ for some $| \psi \rangle \in \mathcal{H} \ox \mathcal{H}$, and the observables are coarse-grained collective observables of the form
\begin{align}
x_A = \sum_{i=1}^{N} \frac{A_1^{(i)} }{\sqrt{N}} \, , \qquad p_A = \sum_{i=1}^{N} \frac{A_2^{(i)} }{\sqrt{N}} \, ,  \nonumber \\
x_B = \sum_{j=1}^{N} \frac{B_1^{(j)} }{\sqrt{N}} \, , \qquad p_B = \sum_{j=1}^{N} \frac{B_2^{(j)}}{\sqrt{N}} \, ,
\end{align}
we have
\begin{align}
    \Var(x_A) & = \langle x_A^2 \rangle - \langle x_A \rangle^2 \nonumber \\
    & = \langle \psi |^{\ox N} \left( \sum_{i=1}^{N} \frac{A_1^{(i)} }{\sqrt{N}} \right)^2  | \psi \rangle^{\ox N} - \left( \langle \psi |^{\ox N} \left( \sum_{i=1}^{N} \frac{A_1^{(i)} }{\sqrt{N}} \right)  | \psi \rangle^{\ox N} \right)^2 \nonumber \\
    & = \frac{1}{N} \sum_{i,j=0}^N \langle \psi |^{\ox N} A_1^{(i)} A_1^{(j)}  | \psi \rangle^{\ox N} -  \left( \frac{1}{\sqrt{N}} \sum_{i=1}^{N} \langle \psi |^{\ox N} A_1^{(i)}  | \psi \rangle^{\ox N} \right)^2 \nonumber \\
    & = \frac{1}{N} \left(  \sum_{i=0}^N \langle \psi |^{\ox N} (A_1^{(i)})^2  | \psi \rangle^{\ox N} + \sum_{i \neq j}^N \langle \psi |^{\ox N} A_1^{(i)} A_1^{(j)}  | \psi \rangle^{\ox N} \right)-  \left( \frac{1}{\sqrt{N}} \sum_{i=1}^{N} \langle \psi | A_1  | \psi \rangle \right)^2 \nonumber \\
    & = \frac{1}{N} \Big( N \langle \psi | A_1^2 | \psi \rangle + N(N-1) \langle \psi | A_1 | \psi \rangle^2 \Big)-  \left( \sqrt{N} \langle \psi | A_1  | \psi \rangle \right)^2 \nonumber \\
    & = \langle \psi | A_1^2   | \psi \rangle - \langle \psi | A_1   | \psi \rangle^2 \nonumber \\
    & = \langle A_1^2  \rangle - \langle A_1    \rangle^2 \nonumber \\
    & = \Var(A_1) \, ,
\end{align}
where now $\langle X \rangle = \langle \psi | X | \psi \rangle$ and $\Var(X) = \langle X^2 \rangle - \langle X \rangle^2$. Similarly, $\Var(x_B)  = \Var( B_1)$, $\Var(p_A) = \Var(A_2 )$ and $\Var(p_B) = \Var( B_2)$. On the other hand,
\begin{align}
    \Cov(x_A,x_B) & = \langle x_A \ox x_B \rangle - \langle x_A \rangle \langle x_B \rangle  \nonumber \\
    & = \langle \psi |^{\ox N} \left( \sum_{i=1}^{N} \frac{A_1^{(i)} }{\sqrt{N}} \right) \ox  \left( \sum_{j=1}^{N} \frac{B_1^{(j)} }{\sqrt{N}} \right)  | \psi \rangle^{\ox N} -  \langle \psi |^{\ox N} \left( \sum_{i=1}^{N} \frac{A_1^{(i)} }{\sqrt{N}} \right)  | \psi \rangle^{\ox N} \, \langle \psi |^{\ox N} \left( \sum_{i=1}^{N} \frac{B_1^{(i)} }{\sqrt{N}} \right)  | \psi \rangle^{\ox N}  \nonumber \\
    & = \frac{1}{N} \sum_{i,j=0}^N \langle \psi |^{\ox N} A_1^{(i)} \ox B_1^{(j)}  | \psi \rangle^{\ox N} -  \left( \frac{1}{\sqrt{N}} \sum_{i=1}^{N} \langle \psi |^{\ox N} A_1^{(i)}  | \psi \rangle^{\ox N} \right) \left( \frac{1}{\sqrt{N}} \sum_{i=1}^{N} \langle \psi |^{\ox N} B_1^{(i)}  | \psi \rangle^{\ox N} \right) \nonumber \\
    & = \frac{1}{N} \Big( N \langle \psi | A_1 \ox B_1  | \psi \rangle + N(N-1) \langle \psi | A_1 | \psi \rangle \langle \psi | B_1  | \psi \rangle \Big)-  \left( \sqrt{N} \langle \psi | A_1  | \psi \rangle \right) \left( \sqrt{N} \langle \psi | B_1  | \psi \rangle \right) \nonumber \\
    & = \langle  A_1 \ox B_1 \rangle - \langle  A_1  \rangle \langle B_1 \rangle \nonumber \\
    & = \Cov(A_1  ,   B_1) \, .
\end{align}
and similarly $\Cov(p_A, p_B) = \Cov(A_2  ,   B_2)$. Finally, 
\begin{align}
    \langle [x_A, p_A] \rangle & = \langle \psi |^{\ox N} \left[ \sum_{i=0}^N \frac{A_1^{(i)}}{\sqrt{N}} ,  \sum_{j=0}^N \frac{A_2^{(j)}}{\sqrt{N}} \right] | \psi \rangle^{\ox N} \nonumber \\
    & = \frac{1}{N} \sum_{i,j=0}^N \langle \psi |^{\ox N} \big[ A_1^{(i)} ,  A_2^{(j)} \big] | \psi \rangle^{\ox N} \nonumber \\
    & =  \frac{1}{N} \sum_{i=0}^N \langle \psi |^{\ox N} \big[ A_1^{(i)} ,  A_2^{(i)} \big] | \psi \rangle^{\ox N} \nonumber \\
    & = \langle \psi | \big[ A_1 ,  A_2 \big]  
 | \psi \rangle \nonumber \\
    & = \langle [ A_1, A_2]   \rangle \, ,
\end{align}
and similarly $\langle [ x_B, p_B] \rangle = \langle [B_1, B_2] \rangle$. Therefore, the inequality reads
\begin{align}
    \Var(A_1 +  B_1) + \Var(A_2 - B_2) \geq \big| \langle [A_1, A_2]   \rangle \big| + \big| \langle  [B_1, B_2] \rangle \big| \, ,
\end{align}
proving Equation \eqref{eq:duansimoniid}.

\section{Robustness of the RME}
\label{app:RME}
\subsection{Decoherence}
\label{app:RMEdecoherence}
Consider the single-particle depolarizing channel
\begin{align}
    \G: \r \mapsto (1-\l) \r + \frac{\l}{d} \,  \mathbb{I}_d \, ,
\end{align}
with $\l \in [0,1]$. For a two-particle state $\r$ we have
\begin{align} 
    \G \ox \G : \r \mapsto 
    (1-\l)^2  \r + 
    \l (1-\l) \Big( \tr_B \r \ox \frac{\mathbb{I}_d}{d}+ \frac{\mathbb{I}_d}{d} \ox \tr_A \r  \Big) 
    + \l^2    \frac{\mathbb{I}_d}{d} \ox  \frac{\mathbb{I}_d}{d}.
\end{align}
For our case where $d=2$, by a similar calculation as before we have
\begin{align}
    \Var(x_A) & = \tr \Big[ (\G \ox \G) ( \r ) A_1^2 \ox \mathbb{I}  \Big] - \tr \Big[ (\G \ox \G) ( \r ) A_1  \ox \mathbb{I} \Big]^2 \nonumber \\
    & = (1-\l) \langle A_1^2 \rangle  + \l  \,  \tr \frac{A_1^2}{2}   -  \left(  (1-\l) \langle A_1  \rangle  + \l \,  \tr \frac{A_1}{2}  \right)^2 \, ,
\end{align}
and analogously for $p_A$, $x_B$ and $p_B$. On the other hand
\begin{align}
    \Cov (x_A, x_B) & = \tr \Big[ (\G \ox \G) ( \r ) A_1 \ox B_1 \Big] -   \tr \Big[ (\G \ox \G) ( \r ) A_1  \Big] \tr \Big[ (\G \ox \G) ( \r ) B_1 \Big]  \nonumber \\ 
    & = (1-\l)^2 \langle A_1 \ox B_1 \rangle + \l (1-\l) \langle A_1  \rangle \, \tr \frac{B_1}{2} + \l (1-\l)  \langle  B_1 \rangle \tr \frac{A_1}{2}+ \l^2 \, \tr \frac{A_1}{2} \, \tr \frac{B_1}{2} \nonumber \\
    & \qquad - \left( (1-\l)  \langle A_1  \rangle  + \l \, \tr \frac{A_1}{2} \right) \left( (1-\l)  \langle B_1 \rangle  + \l \, \tr \frac{B_1}{2} \right) \, 
\end{align}
and analogously for $\Cov(p_A, p_B)$. Finally,
\begin{align}
    \langle [ x_A, p_A ] \rangle & = \tr \Big[ (\G \ox \G) ( \r ) [ A_1, A_2 ]  \Big] \nonumber \\
    & =  (1-\l)  \langle [ A_1, A_2 ]  \rangle  +  \l  \,  \tr \frac{[ A_1, A_2 ]}{2} \, ,
\end{align}
and similarly for $\langle [ x_B, p_B ] \rangle$. Now, for the particular case with the state $\r = \ket{\psi} \bra{\psi}$ with $\ket{\psi} = \cos \frac{\pi}{8} \ket{00} - \sin \frac{\pi}{8} \ket{11}$ and the observables $A_1 = B_1 = \s_x$ and $A_2 = B_2 = \s_y$, we have
\begin{align}
    \Var(x_A) =  \Var(x_B) = \Var(p_A) = \Var(p_B) =  1 \, ,
\end{align}
\begin{align}
    \Cov(x_A, x_B) = - \Cov(p_A,p_B) = - \frac{\sqrt{2}}{2} (1-\l)^2 
\end{align}
and
\begin{align}
    \langle [x_A, p_A ] \rangle = \langle [x_B, p_B ] \rangle = i \sqrt{2} (1-\l) \, .
\end{align}
With these expressions, the inequality is violated if 
\begin{align}
    \l < \frac{3-\sqrt{1+4\sqrt{2}}}{2} \simeq 0.21 \, .
\end{align}

\subsection{Particle losses}
\label{app:RMElosses}
We now compute the maximal loss probability $p$ such that the effective state
\begin{align}
     \r_{2N}^{\text{eff}} = \bigoplus_{M=0}^{2N}  P_{2N}(M) \,  \sum_{\pi \in \mathfrak{S}_{2N}} \frac{\tr_{\pi(1) \dots \pi(2N-M)} \big[ ( \ket{\psi} \bra{\psi} )^{\ox N}
     \big] }{(2N)!} 
\label{eq:app:rhoneff}
\end{align}
gives a violation of the entanglement inequality with observables of the form
\begin{align}
x_A = \sum_{i=1}^{N} \frac{A_1^{(i)} }{\sqrt{N}} \, , \qquad p_A = \sum_{i=1}^{N} \frac{A_2^{(i)} }{\sqrt{N}} \, ,  \nonumber \\
x_B = \sum_{i=1}^{N} \frac{B_1^{(i)} }{\sqrt{N}} \, , \qquad p_B = \sum_{i=1}^{N} \frac{B_2^{(i)}}{\sqrt{N}} \, .
\label{eq:app:4obs}
\end{align}
However, in order to perform this calculation, it turns simpler to account for the effect of particle losses not in the state, as in Equation \eqref{eq:app:rhoneff},  but in the measurements \eqref{eq:app:4obs}. For this purpose, let us introduce a random variable $\a_i$ to each of Alice's particles, where $\a_i=0$ with probability $p$ and $\a_i=1$ with probability $1-p$, and a random variable $\b_i$ to each of Bob's, where $\b_i=0$ with probability $p$ and $\b_i=1$ with probability $1-p$. Then, it can be seen that measuring observables \eqref{eq:app:4obs} on the state $\r_{2N}^{\text{eff}}$ is equivalent to measuring the observables 
\begin{align}
x_A = \sum_{i=1}^{N} \frac{\alpha_i A_1^{(i)}}{\sqrt{N}},  &\qquad \qquad  
p_A = \sum_{i=1}^{N} \frac{\alpha_i A_2^{(i)} }{\sqrt{N}},  \nonumber \\
x_B = \sum_{i=1}^{N}
\frac{\beta_i B_1^{(i)} }{\sqrt{N}},  &\qquad  \qquad 
p_B = \sum_{i=1}^{N} \frac{\beta_i B_2^{(i)}}{\sqrt{N}}  .
\end{align}
on the state $( \ket{\psi} \bra{\psi} )^{\ox N}$. Therefore we have
\begin{align}
    \Var (x_A) & = \langle x_A^2 \rangle - \langle x_A \rangle ^2 \nonumber \\
    & = \E_{\{ \a_i \} } \left[ \langle \psi |^{\ox N} \left( \sum_{i=1}^{N} \frac{\a_i A_1^{(i)} }{\sqrt{N}} \right)^2  | \psi \rangle^{\ox N} \right] - \left( \E_{\{ \a_i \} } \left[   \langle \psi |^{\ox N} \left( \sum_{i=1}^{N} \frac{\a_i A_1^{(i)} }{\sqrt{N}} \right)  | \psi \rangle^{\ox N}  \right] \right)^2 \nonumber \\
    & = \frac{1}{N} \E_{\{ \a_i \} } \left[ \sum_{i,j=1}^N \a_i \a_j  \langle \psi |^{\ox N} A_1^{(i)} A_1^{(j)}  | \psi \rangle^{\ox N} \right] - \left( \frac{1}{\sqrt N} \E_{\{ \a_i \} } \left[ \sum_{i=1}^N \a_i  \langle \psi |^{\ox N} A_1^{(i)}  | \psi \rangle^{\ox N} \right] \right)^2 \nonumber \\
    & = \frac{1}{N} \Big\{ N (1-p) \langle \psi | A_1^2 | \psi \rangle + N(N-1) (1-p)^2  \langle \psi | A_1   | \psi \rangle \langle \psi | A_1   | \psi \rangle \Big\} - \frac{1}{N} \Big( N (1-p)  \langle \psi |A_1  | \psi \rangle \Big)^2 \nonumber \\
    & = (1-p) \langle A_1^2  \rangle - (1-p)^2 \langle A_1 \rangle^2 \, .
\end{align}
Similarly, $\Var(x_B) = (1-p) \langle   B_1^2 \rangle - (1-p)^2 \langle    B_1\rangle^2$, $\Var(p_A) = (1-p) \langle A_2^2 \rangle - (1-p)^2 \langle A_2   \rangle^2$ and $\Var(p_B) = (1-p) \langle    B_2^2 \rangle - (1-p)^2 \langle  B_2\rangle^2$. On the other hand,
\begin{align}
    \Cov(x_A, x_B) & = \langle x_A \ox x_B \rangle - \langle x_A \rangle \langle x_B \rangle \nonumber \\
    & = \E_{\{ \a_i, \b_j \} } \left[ \langle \psi |^{\ox N} \left( \sum_{i=1}^{N} \frac{\a_i A_1^{(i)} }{\sqrt{N}} \right) \ox \left( \sum_{j=1}^{N} \frac{\b_j B_1^{(j)} }{\sqrt{N}} \right)  | \psi \rangle^{\ox N} \right] \nonumber \\
    & \qquad - \left( \E_{\{ \a_i \} } \left[   \langle \psi |^{\ox N} \left( \sum_{i=1}^{N} \frac{\a_i A_1^{(i)} }{\sqrt{N}} \right)  | \psi \rangle^{\ox N}  \right] \right) \left( \E_{\{ \b_j \} } \left[   \langle \psi |^{\ox N} \left( \sum_{j=1}^{N} \frac{\b_j B_1^{(j)} }{\sqrt{N}} \right)  | \psi \rangle^{\ox N}  \right] \right) \nonumber \\
    & = \frac{1}{N} \E_{\{ \a_i, \b_j \} } \left[  \sum_{i,j=1}^{N} \a_i \b_j  \langle \psi |^{\ox N} A_1^{(i)} \ox B_1^{(j)} | \psi \rangle^{\ox N} \right] \nonumber \\
    & \qquad - \frac{1}{N}  \left( \E_{\{ \a_i \} } \left[  \sum_{i=1}^{N} \a_i \langle \psi |^{\ox N} A_1^{(i)}  | \psi \rangle^{\ox N}  \right] \right) \left( \E_{\{ \b_j \} } \left[  \sum_{j=1}^{N} \b_j \langle \psi |^{\ox N} B_1^{(j)}   | \psi \rangle^{\ox N}  \right] \right) \nonumber \\
    & = \frac{1}{N} \Big\{ N (1-p)^2  \langle \psi | A_1 \ox  B_1 | \psi \rangle + N(N-1) (1-p)^2 \langle \psi | A_1 | \psi \rangle \langle \langle \psi |  B_1 | \psi \rangle  \Big\} \nonumber \\
    & \qquad - \frac{1}{N}  \Big( N (1-p)  \langle \psi |A_1  | \psi \rangle \Big) \Big( N(1-p) \langle \psi | B_1 | \psi \rangle \Big) \nonumber \\
    & = (1-p)^2 \langle A_1 \ox B_1 \rangle - (1-p)^2 \langle A_1 \rangle \langle B_1 \rangle \nonumber \\
    & = (1-p)^2 \Cov(A_1, B_1)
\end{align}
and similarly $\Cov(p_A, p_B) = (1-p)^2 \Cov(A_2, B_2)$. Finally, \begin{align}
    \langle [x_A, p_A] \rangle & =   \E_{\{ \a_i \} } \left[ \langle \psi |^{\ox N} \left[ \sum_{i=0}^N \frac{\a_i A_1^{(i)}}{\sqrt{N}} ,  \sum_{j=0}^N \frac{ \a_j A_2^{(j)}}{\sqrt{N}} \right] | \psi \rangle^{\ox N} \right] \nonumber \\
    & = \frac{1}{N} \E_{\{ \a_i \} } \left[ \sum_{i,j=0}^N \a_i \a_j  \langle \psi |^{\ox N} \big[ A_1^{(i)} ,  A_2^{(j)} \big] | \psi \rangle^{\ox N} \right] \nonumber \\
    & = \frac{1}{N} \E_{\{ \a_i \} } \left[ \sum_{i=0}^N \a_i^2  \langle \psi |^{\ox N} \big[ A_1^{(i)} ,  A_2^{(i)} \big] | \psi \rangle^{\ox N} \right] \nonumber \\
    & =  \frac{1}{N} \Big( N (1-p) \langle \psi | \big[ A_1,  A_2 \big] | \psi \rangle \Big) \nonumber \\
    & = (1-p) \langle [ A_1, A_2] \rangle \, ,
\end{align}
and similarly $\langle [ x_B, p_B] \rangle = (1-p) \langle [B_1, B_2] \rangle$. For the particular case with the state $\r = \ket{\psi} \bra{\psi}$ with $\ket{\psi} = \cos \frac{\pi}{8} \ket{00} - \sin \frac{\pi}{8} \ket{11}$ and the observables $A_1 = B_1 = \s_x$ and $A_2 = B_2 = \s_y$, we have
\begin{align}
    \Var(x_A) = \Var(p_A) = \Var(x_B) = \Var(p_B) =  1-p  \, ,
\end{align}
\begin{align}
    \Cov(x_A, x_B) = - \Cov(p_A,p_B) = - \frac{\sqrt
    {2}}{2} (1-p)^2 
\end{align}
and
\begin{align}
    \langle [x_A, p_A ] \rangle = \langle [x_B, p_B ] \rangle = i \sqrt{2} (1-p)   \, .
\end{align}
With these expressions, the inequality is violated if 
\begin{align}
    p < 2-\sqrt{2} \simeq 0.59 \, .
\end{align}

\subsection{Non-projective measurements}
Let us now suppose that instead of the projective measurements $A_1$, $A_2$, $B_1$ and $B_2$ that lead to the maximal violation of the inequality ($A_1 = B_1=\s_x$, $A_2=B_2=\s_y$), some POVMs with elements $\{E_1^{(1)},E_1^{(-1)} \}$, $\{E_2^{(1)},E_2^{(-1)} \}$, $\{F_1^{(1)},F_1^{(-1)} \}$ and $\{F_2^{(1)},F_2^{(-1)} \}$ are actually measured. Let us also suppose that instead of the theoretical commutators $[A_1, A_2]= 2 i \s_z =: 2 i A_3$ and $[B_1, B_2] =: 2 i \s_z =: 2 i B_3$, some POVMs with elements $\{E_3^{(1)},E_3^{(-1)} \}$ and $\{F_3^{(1)},F_3^{(-1)} \}$ are actually measured. For this, we write the projective decomposition of the single-particle observables 
\begin{align}
    A_j = \sum_{a \in \{1,-1\}} a \, P_j^{(a)} = P_j^{(1)}-P_j^{(-1)} \, , \quad  B_j = \sum_{b \in \{1,-1\}} b \, Q_j^{(b)} = Q_j^{(1)}-Q_j^{(-1)}\, , \quad j \in \{1, 2, 3 \} \, ,
\end{align}
where $P_j^{(a)}$ ($Q_j^{(b)}$) is the projector onto the subspace associated with eigenvalue $a$ ($b$) for observable $A_j$ ($B_j$). Then, we let the POVM elements be $\e$-close to the corresponding projectors, that is,
\begin{align}
    E_j^{(a)} = P_j^{(a)} + \e C_j^{(a)} \, , \quad
    F_j^{(b)} = Q_j^{(b)} + \e D_j^{(b)}  \, ,
\end{align}
where $\e > 0$ and $C_j^{(a)}$ and $D_j^{(b)}$ have norm 1. Completeness of the POVMs implies
\begin{align}
    C_j^{(1)} = - C_j^{(-1)} =: C_j \, , \quad D_j^{(1)} = - D_j^{(-1)} =: D_j \, .
\end{align}
Hermiticity of the POVM elements implies hermiticity of the $C_j$'s and $D_j$'s. Finally we have positive semi-definiteness of the POVM elements: 
\begin{align}
    E_j^{(a)} & = P_j^{(a)} + \e a C_j \geq 0 \, , \nonumber \\
    F_j^{(b)} & = Q_j^{(a)} + \e b D_j \geq 0 \, .
\end{align}
With this we have
\begin{align}
    \var(x_A) &  = \avg{x_A^2} - \avg{x_A}^2 \nonumber \\
    & = \left\langle \left( \frac{\sum_{i=1}^N a_i }{\sqrt{N}} \right)^2 \right\rangle - \left\langle \left( \frac{\sum_i a_i }{\sqrt{N}} \right) \right\rangle^2 \nonumber \\
    & = \frac{1}{N} \sum_{i,j=1}^N \avg{a_i a_j} - \frac{1}{N}  \left( \sum_{i=1}^N \avg{a_i} \right)^2 \nonumber \\
    & =  \frac{1}{N} \left\{ \sum_{i=1}^N \avg{a_i^2} + \sum_{i\neq j} \avg{a_i a_j} \right\} - \frac{1}{N} \left( \sum_{i=1}^N \avg{a_i} \right)^2 \nonumber \\
    & = \avg{a^2} + (N-1) \avg{a_1 a_2} - N \avg{a}^2 \nonumber \\
    & = \avg{a^2} - \avg{a}^2 \nonumber \\
    & = \sum_a a^2 \, \tr \r E_1^{(a)} - \left(\sum_a a \, \tr \r E_1^{(a)}  \right)^2 \nonumber \\
    & = \tr \r ( E_1^{(1)} + E_1^{(-1)} ) - \left(\tr \r (E_1^{(1)}-E_1^{(-1)} )  \right)^2 \nonumber \\
    & = \tr \r ( P_1^{(1)} + \e C_1 + P_1^{(-1)} - \e C_1) - \left(\tr \r (P_1^{(1)} + \e C_1 - P_1^{(-1)} + \e C_1)  \right)^2 \nonumber \\
    & = \tr \r A_1^2 - \left(\tr \r (A_1 + 2\e C_1)  \right)^2 \nonumber \\
    & = \avg{A_1^2} - (\avg{A_1} + 2 \e \avg{C_1} )^2 \nonumber \\
    & = \var(A_1) - 4 \e \avg{A_1} \avg{C_1} - 4 \e^2 \avg{C_1}^2 \, .
\end{align}
Similarly,
\begin{align}
    \var(x_B) & =  \var(B_1) - 4 \e \avg{B_1} \avg{D_1} - 4 \e^2 \avg{D_1}^2 \, , \nonumber \\
    \var(p_A) & =   \var(A_2) - 4 \e \avg{A_2} \avg{C_2} - 4 \e^2 \avg{C_2}^2 \, , \nonumber \\
    \var(p_B) & =  \var(B_2) - 4 \e \avg{B_2} \avg{D_2} - 4 \e^2 \avg{D_2}^2 \, .
\end{align}
On the other hand
\begin{align}
    \cov(x_A, x_B) &  = \avg{x_A  x_B} - \avg{x_A} \avg{x_B} \nonumber \\
    & = \left\langle \left( \frac{\sum_{i=1}^N a_i }{\sqrt{N}} \right)  \left( \frac{\sum_{i=1}^N b_i }{\sqrt{N}} \right) \right\rangle - \left\langle \left( \frac{\sum_i a_i }{\sqrt{N}} \right) \right\rangle \left\langle \left( \frac{\sum_i b_i }{\sqrt{N}} \right) \right\rangle \nonumber \\
    & = \frac{1}{N} \sum_{i,j=1}^N \avg{a_i b_j} - \frac{1}{N}  \left( \sum_{i=1}^N \avg{a_i} \right) \left( \sum_{i=1}^N \avg{b_i} \right) \nonumber \\
    & =  \frac{1}{N} \left\{ \sum_{i=1}^N \avg{a_i b_i} + \sum_{i\neq j} \avg{a_i b_j} \right\} - \frac{1}{N} \left( \sum_{i=1}^N \avg{a_i} \right) \left( \sum_{i=1}^N \avg{b_i} \right) \nonumber \\
    & = \avg{ab} + (N-1) \avg{a_1 b_2} - N \avg{a} \avg{b} \nonumber \\
    & = \avg{ab} - \avg{a} \avg{b} \nonumber \\
    & = \sum_{a,b} ab \, \tr \r E_1^{(a)} \ox F_1^{(b)} - \left(\sum_a a \, \tr \r E_1^{(a)}  \right)\left(\sum_b b \, \tr \r F_1^{(b)}  \right) \nonumber \\
    & = \tr \r ( E_1^{(1)} - E_1^{(-1)}) \ox (F_1^{(1)} - F_1^{(-1)}) - \left(\tr \r (E_1^{(1)}-E_1^{(-1)})  \right) \left(\tr \r (F_1^{(1)}-F_1^{(-1)})  \right) \nonumber \\
    & = \tr \r ( P_1^{(1)} + \e C_1 - P_1^{(-1)} + \e C_1) \ox ( Q_1^{(1)} + \e D_1 - Q_1^{(-1)} + \e D_1) \nonumber \\
    & \qquad - \left(\tr \r (P_1^{(1)} + \e C_1 - P_1^{(-1)} + \e C_1)  \right) \left(\tr \r (Q_1^{(1)} + \e D_1 - Q_1^{(-1)} + \e D_1)  \right) \nonumber \\
    & = \tr \r (A_1 + 2 \e C_1 )  \ox ( B_1 + 2 \e D_1 ) - \left(\tr \r (A_1 + 2\e C_1)  \right) \left(\tr \r (B_1 + 2\e D_1)  \right) \nonumber \\
    & = \avg{A_1\ox B_1} + 2 \e \avg{A_1 \ox D_1} + 2 \e \avg{C_1 \ox B_1} + 4 \e^2 \avg{C_1 \ox D_1} - (\avg{A_1} + 2 \e \avg{C_1} ) (\avg{B_1} + 2 \e \avg{D_1} ) \nonumber \\
    & = \cov(A_1, B_1) +2 \e \, \cov(A_1, D_1)  + 2 \e \, \cov(C_1,B_1) + 4 \e^2 \cov(C_1,D_1) \, .
\end{align}
Similarly, 
\begin{align}
    \cov(p_A, p_B) = \cov(A_2, B_2) +2 \e \, \cov(A_2, D_2)  + 2 \e \, \cov(C_2,B_2) + 4 \e^2 \cov(C_2,D_2) \, .
\end{align}
Finally, the commutators are 
\begin{align}
    \avg{[x_A, p_A]} = 2 i ( \avg{A_3} + 2 \e \avg{C_3} ) \, , \nonumber \\
    \avg{[x_B, p_B]} = 2 i (\avg{B_3} + 2 \e \avg{D_3} ) \, .
\end{align}
For the particular case with the state $\r = \ket{\psi} \bra{\psi}$ with $\ket{\psi} = \cos \frac{\pi}{8} \ket{00} - \sin \frac{\pi}{8} \ket{11}$ and the observables $A_1 = B_1 = \s_x$ and $A_2 = B_2 = \s_y$, we have
\begin{align}
    \var(x_A) & = 1 - 4 \e^2 \avg{C_1}^2 \, , \nonumber \\
    \var(x_B) & = 1 - 4 \e^2 \avg{D_1}^2 \, , \nonumber \\
    \var(p_A) & = 1 - 4 \e^2 \avg{C_2}^2 \, , \nonumber \\
    \var(p_B) & = 1 - 4 \e^2 \avg{D_2}^2 \, ,
\end{align}
and
\begin{align}
    \cov(x_A, x_B) & = - \frac{1}{\sqrt{2}} + 2 \e \avg{A_1 \ox D_1} + 2 \e \avg{C_1 \ox B_1} +4 \e^2 \cov(C_1, D_1) \, , \nonumber \\
    \cov(p_A, p_B) & =   \frac{1}{\sqrt{2}} + 2 \e \avg{A_2 \ox D_2} + 2 \e \avg{C_2 \ox B_2} +4 \e^2 \cov(C_2, D_2) \, , 
\end{align}
and
\begin{align}
    \avg{[x_A, p_A]} & = i\sqrt{2} + 4 i \e \avg{C_3} \, , \nonumber \\
    \avg{[x_B, p_B]} & = i\sqrt{2} + 4 i \e \avg{D_3} \, .
\end{align}
Optimizing for the wirst case scenario with the $C_j$'s and $D_j$'s Hermitian norm-1 operators, we find that the violation of the inequality is still visible as long as
\begin{align}
    \e < \frac{\sqrt{2}-1}{3 \sqrt{2}} \simeq 0.10 \, .
\end{align}

\section{Inequality for a $q$-bipartition of IID States}
\label{app:duanrandombipartition}
For a $q$-bipartition of the $2N$ particle system described by Equation \eqref{eq:rho2nq} we have
\begin{align}
    \avg{x_A} & = \tr \left( \r_{2N}(q) \sum_{i=1}^{2 N_A + N_{AB}} \frac{A_1^{(i)}}{\sqrt{N}} \right) \nonumber \\
    & = \frac{1}{\sqrt{N}} \sum_{N_A=0}^N \sum_{N_B=0}^{N-N_A} P_N(q, N_A, N_B)  \sum_{i=1}^{2 N_A + N_{AB}}  \tr \Big( \s_{A A}^{\otimes N_A} \ox \s_{A B}^{\otimes N_{AB}} \ox \s_{B B}^{\otimes N_B} A_1^{(i)} \Big) \nonumber \\ 
    & = \frac{1}{\sqrt{N}} \sum_{N_A=0}^N \sum_{N_B=0}^{N-N_A} P_N(q, N_A, N_B) (2 N_A + N_{AB} ) \avg{A_1} \nonumber \\
    & = \frac{1}{\sqrt{N}} 2 N q \avg{A_1} \nonumber \\
    & = 2 q \sqrt{N} \avg{A_1} \, ,
\end{align}
where $\langle A_1 \rangle = \tr \big[ \s A_1 \big]$, and
\begin{align}
    \avg{x_A^2} & = \tr \left( \r_{2N}(q) \left(\sum_{i=1}^{2 N_A + N_{AB}} \frac{A_1^{(i)}}{\sqrt{N}} \right)^2 \right) \nonumber \\
    & = \frac{1}{N} \sum_{N_A=0}^N \sum_{N_B=0}^{N-N_A} P_N(q, N_A, N_B)  \sum_{i,j=1}^{2 N_A + N_{AB}}  \tr \Big( \s_{A A}^{\otimes N_A} \ox \s_{A B}^{\otimes N_{AB}} \ox \s_{B B}^{\otimes N_B} A_1^{(i)} A_1^{(j)} \Big) \nonumber \\ 
    & = \frac{1}{N} \sum_{N_A=0}^N \sum_{N_B=0}^{N-N_A} P_N(q, N_A, N_B)  \sum_{i,j=1}^{2 N_A + N_{AB}}  \tr \Big( \s_{A A}^{\otimes N_A} \ox \s_{A B}^{\otimes N_{AB}} A_1^{(i)} A_1^{(j)} \Big) \nonumber \\ 
    & = \frac{1}{N} \sum_{N_A=0}^N \sum_{N_B=0}^{N-N_A} P_N(q, N_A, N_B) \Bigg\{   \sum_{i,j=1}^{2 N_A }  \tr \Big( \s_{A A}^{\otimes N_A}  A_1^{(i)} A_1^{(j)}  \Big) + 2  \sum_{i=1}^{2 N_A } \sum_{j=2N_A+1}^{2 N_A + N_{AB}} \tr \Big( \s_{A A}^{\otimes N_A} \ox \s_{A B}^{\otimes N_{AB}} A_1^{(i)} A_1^{(j)}  \Big)  \nonumber \\
    & \qquad + \sum_{i,j=2N_A+1}^{2 N_A + N_{AB}}  \tr \Big(  \s_{A B}^{\otimes N_{AB}}  A_1^{(i)} A_1^{(j)} \Big) \Bigg\} \nonumber \\ 
    & = \frac{1}{N} \sum_{N_A=0}^N \sum_{N_B=0}^{N-N_A} P_N(q, N_A, N_B) \Bigg\{ \sum_{i=1}^{2 N_A }  \tr \Big( \s_{A A}^{\otimes N_A}    A_1^{(i)} A_1^{(i)}  \Big) + \sum_{i=1}^{2 N_A }  \tr \Big( \s_{A A}^{\otimes N_A}    A_1^{(i)} A_1^{(i-(-1)^i)}  \Big)  + \sum_{\substack{i,j=1 \\ j \neq i \\ j \neq i - (-1)^i}}^{2 N_A }  \tr \Big( \s_{A A}^{\otimes N_A}    A_1^{(i)} A_1^{(j)}  \Big)  \nonumber \\
    & \qquad  + 2 (2 N_A) N_{AB} \avg{A_1}^2 + \sum_{i=2N_A +1}^{2 N_A + N_{AB}}  \tr \Big(   \s_{A B}^{\otimes N_{AB}}  A_1^{(i)} A_1^{(i)} \Big)  +  \sum_{\substack{i,j=2 N_A +1 \\ j \neq i}}^{2 N_A + N_{AB}}  \tr \Big(  \s_{A B}^{\otimes N_{AB}}  A_1^{(i)} A_1^{(j)} \Big) \Bigg\} \nonumber \\
    & = \frac{1}{N} \sum_{N_A=0}^N \sum_{N_B=0}^{N-N_A} P_N(q, N_A, N_B) \bigg\{ 2 N_A   \avg{   A_1^2} + 2 N_A \avg{A_1 \ox A_1 } + 2 N_A (2 N_A -2) \avg{A_1}^2 + 2 (2 N_A) N_{AB} \avg{A_1}^2 \nonumber \\
    & \qquad \qquad \qquad \qquad \qquad \qquad \qquad \qquad  + N_{AB} \avg{A_1^2}  + N_{AB} (N_{AB} -1) \avg{A_1}^2 \bigg\} \nonumber \\
    & = \frac{1}{N} \sum_{N_A=0}^N \sum_{N_B=0}^{N-N_A} P_N(q, N_A, N_B) \bigg\{ (2 N_A + N_{AB} ) \avg{A_1^2} + 2 N_A \avg{A_1 \ox A_1} \nonumber \\
    & \qquad \qquad \qquad \qquad \qquad \qquad \qquad \qquad  + \Big( 2 N_A (2 N_A -2) + 2 (2 N_A) N_{AB} + N_{AB} (N_{AB} -1) \Big) \avg{A_1}^2 \bigg\} \nonumber \\
    & = 2 q \avg{A_1^2} + 2 q^2 \avg{A_1 \ox A_1} + 4 (N-1) q^2 \avg{A_1}^2 \, ,
\end{align}
so that 
\begin{align}
    \Var(x_A) = 2 q \avg{A_1^2} + 2 q^2 \avg{A_1 \ox A_1} - 4 q^2 \avg{A_1}^2 \, .
\end{align}
Similarly
\begin{align}
    \Var(x_B) & = 2 (1-q) \avg{B_1^2} + 2 (1-q)^2 \avg{B_1 \ox B_1} - 4 (1-q)^2 \avg{B_1}^2 \, ,\nonumber \\
    \Var(p_A) & = 2 q \avg{A_2^2} + 2 q^2 \avg{A_2 \ox A_2} - 4 q^2 \avg{A_2}^2 \, ,\nonumber \\
    \Var(p_B) & = 2 (1-q) \avg{B_2^2} + 2 (1-q)^2 \avg{B_2 \ox B_2} - 4 (1-q)^2 \avg{B_2}^2 \, .
\end{align}
On the other hand, 
\begin{align}
    \avg{x_A \ox x_B} & = \tr \left( \r_{2N}(q) \left( \sum_{i=1}^{2 N_A + N_{AB}} \frac{A_1^{(i)}}{\sqrt{N}} \right)  \left(\sum_{j=1}^{2 N_B + N_{AB}} \frac{B_1^{(j)}}{\sqrt{N}} \right) \right) \nonumber \\
    & = \frac{1}{N} \sum_{N_A=0}^N \sum_{N_B=0}^{N-N_A} P_N(q, N_A, N_B)  \sum_{i=1}^{2 N_A + N_{AB}} \sum_{j=1}^{2 N_B + N_{AB}}  \tr \Big( \s_{A A}^{\otimes N_A} \ox \s_{A B}^{\otimes N_{AB}} \ox \s_{B B}^{\otimes N_B} A_1^{(i)} B_1^{(j)} \Big) \nonumber \\ 
    & = \frac{1}{N} \sum_{N_A=0}^N \sum_{N_B=0}^{N-N_A} P_N(q, N_A, N_B)  \Bigg\{ \sum_{i=1}^{2N_A} \sum_{j=1}^{2N_B} \tr \Big( \s_{A A}^{\otimes N_A} \ox \s_{A B}^{\otimes N_{AB}} \ox \s_{B B}^{\otimes N_B} A_1^{(i)} B_1^{(j)} \Big)  \nonumber \\
    & \qquad \qquad  +  \sum_{i=2N_A+1}^{2N_A+N_{AB}} \sum_{j=1}^{2N_B} \tr \Big( \s_{A A}^{\otimes N_A} \ox \s_{A B}^{\otimes N_{AB}} \ox \s_{B B}^{\otimes N_B} A_1^{(i)} B_1^{(j)} \Big) \nonumber \\
    & \qquad \qquad  + \sum_{i=2N_A+1}^{2N_A+N_{AB}} \sum_{j=1}^{2N_B} \tr \Big( \s_{A A}^{\otimes N_A} \ox \s_{A B}^{\otimes N_{AB}} \ox \s_{B B}^{\otimes N_B} A_1^{(i)} B_1^{(j)} \Big) \nonumber \\
    & \qquad \qquad  +  \sum_{i=2N_A+1}^{2N_A+N_{AB}} \sum_{j=2N_B+1}^{2N_B+N_{AB}} \tr \Big( \s_{A A}^{\otimes N_A} \ox  \s_{A B}^{\otimes N_{AB}} \ox \s_{B B}^{\otimes N_B} A_1^{(i)} B_1^{(j)} \Big) \Bigg\} \nonumber \\
    & = \frac{1}{N} \sum_{N_A=0}^N \sum_{N_B=0}^{N-N_A} P_N(q, N_A, N_B)  \bigg\{ (2 N_A)(2 N_B) \avg{A_1} \avg{B_1}   +  2N_A N_{AB} \avg{A_1} \avg{B_1} + N_{AB} 2 N_B \avg{A_1} \avg{B_1} \nonumber \\
    & \qquad +  N_{AB} \avg{A_1 \ox B_1} + N_{AB} (N_{AB}-1) \avg{A_1} \avg{B_1} \bigg\} \nonumber \\
    & =\frac{1}{N} \sum_{N_A=0}^N \sum_{N_B=0}^{N-N_A} P_N(q, N_A, N_B)  \bigg\{ \Big( (2N_A +N_{AB})(2N_B + N_{AB})-N_{AB} \Big) \avg{A_1} \avg{B_1} + N_{AB} \avg{A_1 \ox B_1} \bigg\} \nonumber \\
    & = 4 (N-1) q (1-q) \avg{A_1}\avg{B_1} + 2q(1-q) \avg{A_1 \ox B_1} \, ,
\end{align}
so that
\begin{align}
    \Cov(x_A, x_B) = 2 q (1-q) \avg{A_1 \ox B_1} - 4 q (1-q) \avg{A_1} \avg{B_1} \, .
\end{align}
Similarly, 
\begin{align}
    \Cov(p_A, p_B) = 2 q (1-q) \avg{A_2 \ox B_2} - 4 q (1-q) \avg{A_2} \avg{B_2} \, .
\end{align}
Finally, 
\begin{align}
    \avg{[x_A,p_A]} & =  \tr \left( \r_{2N}(q) \left[ \sum_{i=1}^{2 N_A + N_{AB}} \frac{A_1^{(i)}}{\sqrt{N}} \, , \sum_{j=1}^{2 N_B + N_{AB}} \frac{A_2^{(j)}}{\sqrt{N}} \right] \right) \nonumber \\
    & = \frac{1}{N} \sum_{N_A=0}^N \sum_{N_B=0}^{N-N_A} P_N(q, N_A, N_B)  \sum_{i,j=1}^{2 N_A + N_{AB}}  \tr \Big( \s_{A A}^{\otimes N_A} \ox \s_{A B}^{\otimes N_{AB}} \ox \s_{B B}^{\otimes N_B} [A_1^{(i)}, A_2^{(j)}] \Big) \nonumber \\
    & = \frac{1}{N} \sum_{N_A=0}^N \sum_{N_B=0}^{N-N_A} P_N(q, N_A, N_B)  \sum_{i=1}^{2N_A + N_{AB}}   \tr \Big( \s_{A A}^{\otimes N_A} \ox  \s_{A B}^{\otimes N_{AB}} [A_1^{(i)}, A_2^{(i)}] \Big) \nonumber \\
    & = \frac{1}{N} \sum_{N_A=0}^N \sum_{N_B=0}^{N-N_A} P_N(q, N_A, N_B) (2N_A+N_{AB})  \avg{[A_1, A_2]}  \nonumber \\
    & = 2 q \avg{[A_1, A_2]} \, .
\end{align}
Similarly, 
\begin{align}
    \avg{[x_B,p_B]} = 2 (1-q) \avg{[B_1, B_2]} \, .
\end{align}
Therefore, the inequality reads
\begin{align}
     & 2 q \Big( \avg{A_1^2} + q \avg{A_1 \ox A_1} - 2 q \avg{A_1}^2 \Big)  \nonumber \\
     & + 2 (1-q) \Big( \avg{B_1^2} + (1-q) \avg{B_1 \ox B_1} -2 (1-q) \avg{B_1}^2 \Big) \nonumber \\
     & + 4 q (1-q) \Big(  \avg{A_1 \ox B_1} - 2 \avg{A_1} \avg{B_1} \Big) \nonumber \\
     & +  2 q \Big( \avg{A_2^2} + q \avg{A_2 \ox A_2} - 2 q \avg{A_2}^2 \Big)  \nonumber \\
     & + 2 (1-q) \Big( \avg{B_2^2} + (1-q) \avg{B_2 \ox B_2} - 2 (1-q) \avg{B_2}^2 \Big) \nonumber \\
     & - 4q (1-q) \Big( \avg{A_2 \ox B_2} -2 \avg{A_2} \avg{B_2} \Big) \geq \nonumber \\
     & \geq 2 q |\avg{[A_1, A_2]}| + 2 (1-q) |\avg{[B_1, B_2]}| \, ,
\end{align}
proving Equation \eqref{eq:fq} in the main text.

\section{Robustness of the IME}
\label{app:IMErobustness}

\subsection{Decoherence}
\label{app:IMEdecoherence}
We consider the effect of the single-particle depolarizing channel on our two-qutrit state:
\begin{align}
    \s \mapsto (\G \ox \G) (\s) = (1-\l)^2 \s + \l (1-\l) \Big( \tr_B \s \ox \frac{\mathbb{I}_3}{3} + \frac{\mathbb{I}_3}{3} \ox \tr_A \s \Big) + \l^2 \frac{\mathbb{I}_3}{3} \ox \frac{\mathbb{I}_3}{3} \, . 
\end{align}
By a calculation analogous to the one in Appendix \ref{app:duanrandombipartition}, on average, for $q$, we have
\begin{align}
    \Var(x_A) & = \, \tr \Big[ (\G \ox \G )(\s) A_1^2 \Big] + \frac{2}{3} \, \tr \Big[ (\G \ox \G )(\s) A_1 \ox A_1 \Big]- \frac{4}{3} \, \tr \Big[ (\G \ox \G )(\s) A_1 \Big]^2 \nonumber \\
    & =(1-\l)  \avg{A_1^2} + \l \, \tr \frac{A_1^2}{3} \nonumber \\
    & \quad + \frac{2}{3} \left( (1-\l)^2 \avg{A_1 \ox A_1} + 2 \l (1-\l) \avg{A_1} \tr \frac{A_1}{3} + \l^2 \left( \tr\frac{A_1}{3} \right)^2 \right) \nonumber \\
    & \quad - \frac{4}{3} \left( (1-\l)  \avg{A_1} + \l \, \tr \frac{A_1}{3}  \right)^2 \, ,
\end{align}
where $\avg{A_1} = \tr \big[ \s A_1 \big]$. Similarly, 
\begin{align}
    \Var(x_B) & = (1-\l)  \avg{B_1^2} + \l \, \tr \frac{B_1^2}{3}  \nonumber \\
    & \quad + \frac{2}{3} \left( (1-\l)^2 \avg{B_1 \ox B_1} + 2 \l (1-\l) \avg{B_1} \tr \frac{B_1}{3} + \l^2 \left( \tr\frac{B_1}{3} \right)^2 \right) \nonumber \\
    & \quad - \frac{4}{3} \left( (1-\l)  \avg{B_1} + \l \, \tr \frac{B_1}{3}  \right)^2 \, , \nonumber \\
    \Var(p_A) & =   (1-\l)  \avg{A_2^2} + \l \, \tr \frac{A_2^2}{3}  \nonumber \\
    & \quad + \frac{2}{3} \left( (1-\l)^2 \avg{A_2 \ox A_2} + 2 \l (1-\l) \avg{A_2} \tr \frac{A_2}{3} + \l^2 \left( \tr\frac{A_2}{3} \right)^2 \right) \nonumber \\
    & \quad - \frac{4}{3} \left( (1-\l)  \avg{A_2} + \l \, \tr \frac{A_2}{3}  \right)^2 \, , \nonumber \\    
    \Var(p_B) & =  (1-\l)  \avg{B_2^2} + \l \, \tr \frac{B_2^2}{3} \nonumber \\
    & \quad + \frac{2}{3} \left( (1-\l)^2 \avg{B_2 \ox B_2} + 2 \l (1-\l) \avg{B_2} \tr \frac{B_2}{3} + \l^2 \left( \tr\frac{B_2}{3} \right)^2 \right) \nonumber \\
    & \quad - \frac{4}{3} \left( (1-\l)  \avg{B_2} + \l \, \tr \frac{B_2}{3}  \right)^2 \, ,
\end{align}
On the other hand,
\begin{align}
    \Cov(x_A, x_B) & = \frac{1}{3} \left( (1-\l)^2 \avg{A_1 \ox B_1} + \l (1-\l) \avg{A_1} \tr \frac{B_1}{3} + \l (1-\l) \avg{B_1} \tr \frac{A_1}{3} + \l^2 \tr \frac{A_1}{3} \tr \frac{B_1}{3} \right)  \nonumber \\
    & \quad - \frac{2}{3} \left( (1-\l) \avg{A_1} + \l \, \tr \frac{A_1}{3} \right) \left( (1-\l) \avg{B_1} + \l \, \tr \frac{B_1}{3} \right)  \, , \nonumber \\
    \Cov (p_A, p_B) & = \frac{1}{3} \left( (1-\l)^2 \avg{A_2 \ox B_2} + \l (1-\l) \avg{A_2} \tr \frac{B_2}{3} + \l (1-\l) \avg{B_2} \tr \frac{A_2}{3} + \l^2 \tr \frac{A_2}{3} \tr \frac{B_2}{3} \right)  \nonumber \\
    & \quad - \frac{2}{3} \left( (1-\l) \avg{A_2} + \l \, \tr \frac{A_2}{3} \right) \left( (1-\l) \avg{B_2} + \l \, \tr \frac{B_2}{3} \right)  \, .
\end{align}
Finally, 
\begin{align}
    \avg{[x_A,p_A]} & =  (1-\l) \avg{[A_1,A_2]} + \l \, \tr \frac{[A_1,A_2]}{3}  \, , \nonumber \\
    \avg{[x_B,p_B]} & =  (1-\l) \avg{[B_1,B_2]} + \l \, \tr \frac{[B_1,B_2]}{3}   \, .
\end{align}
Then for a state $\ket{\psi} = \sum_{i,j=0}^2 c_{ij} \ket{ij}$ with coefficients as in \eqref{eq:cijnumerical} and the observables defined by angles \eqref{eq:anglesnumerical} we have
\begin{align}
    \Var (x_A) & = \Var (p_B) = \frac{2}{3} + (1-\l) \frac{-1 + 3 c_{11}^2 + 6 c_{02}  \re (c_{00} +  c_{22})}{6} +  (1-\l)^2 \frac{  4 c_{02} c_{11} + 2 c_{11} \re (c_{00} + c_{22} ) }{3} \, , \nonumber \\
    \Var(x_B) & = \Var (p_A) = \frac{2}{3} + (1-\l) \frac{-1 + 3 c_{11}^2 + 6 c_{02}  \big(  \re (c_{00} +  c_{22}) \cos (2 \j ) - \im (c_{00}-c_{22}) \sin (2 \j) \big)}{6} \nonumber \\
    & \qquad \qquad + (1-\l)^2 \frac{  4 c_{02} c_{11} + 2 c_{11} \big( \re (c_{00} + c_{22} ) \cos (2 \j) - \im (c_{00} - c_{22}) \sin (2 \j) \big) }{3} \, , \nonumber \\
    \Cov (x_A, x_B) & = - \Cov(p_A, p_B)  =  (1-\l)^2 \frac{2 c_{11} c_{02} \cos (\j) + c_{11} \big( \re ( c_{00}+c_{22} ) \cos (\j) - \im (c_{00} - c_{22} )  \sin ( \j) \big) }{3} \nonumber \\
    \avg{[x_A, x_B]} & = \avg{[x_B,p_B]} = i (1-\l)  (|c_{00}|^2 -|c_{22}|^2) \sin (\j) \, ,
\end{align}
and the inequality is violated as long as
\begin{align}
    \l \lesssim 0.06 \, .
\end{align}

\subsection{Particle losses}
\label{app:IMElosses}
We model losses as in Appendix \ref{app:RMElosses}, i.e. we consider the observables
\begin{align}
x_A = \sum_{i=1}^{2 N_A + N_{AB}} \frac{\alpha_i A_1^{(i)}}{\sqrt{N}},  &\qquad \qquad  
p_A = \sum_{i=1}^{2N_A + N_{AB}} \frac{\alpha_i A_2^{(i)} }{\sqrt{N}},  \nonumber \\
x_B = \sum_{i=1}^{2 N_B + N_{AB}}
\frac{\beta_i B_1^{(i)} }{\sqrt{N}},  &\qquad  \qquad 
p_B = \sum_{i=1}^{2N_B + N_{AB}} \frac{\beta_i B_2^{(i)}}{\sqrt{N}}  .
\end{align}
where each $\a_i$ and $\b_i$ are $0$ with probability $p$ and are $1$ with probability $1-p$. For a $q$-bipartition we then have
\begin{align}
    \Var(x_A) & =  (1-p) \avg{A_1^2} + \frac{2}{3} (1-p)^2 \avg{A_1 \ox A_1} -\frac{4}{3} (1-p)^2 \avg{A_1}^2 \, , \nonumber \\
    \Var(x_B) & = (1-p) \avg{B_1^2} +\frac{2}{3} (1-p)^2 \avg{B_1 \ox B_1} -\frac{4}{3} (1-p)^2 \avg{B_1}^2 \, ,\nonumber \\
    \Var(p_A) & = (1-p) \avg{A_2^2} + \frac{2}{3} (1-p)^2 \avg{A_2 \ox A_2} - \frac{4}{3} (1-p)^2 \avg{A_2}^2 \, ,\nonumber \\
    \Var(p_B) & =  (1-p)\avg{B_2^2} +\frac{2}{3} (1-p)^2 \avg{B_2 \ox B_2} - \frac{4}{3}(1-p)^2 \avg{B_2}^2 \, , \nonumber \\
    \Cov(x_A, x_B) & = (1-p)^2 \Big( \frac{1}{3} \avg{A_1 \ox B_1} - \frac{2}{3} \avg{A_1} \avg{B_1} \Big) \, , \nonumber \\
    \Cov(p_A, p_B) & =  (1-p)^2 \Big( \frac{1}{3} \avg{A_2 \ox B_2} -\frac{2}{3} \avg{A_2} \avg{B_2} \Big) \, , \nonumber \\
    \avg{[x_A,p_A]} & =  (1-p) \avg{[A_1,A_2]} \, , \nonumber \\
    \avg{[x_B,p_B]} & =   (1-p) \avg{[B_1,B_2]} \, .
\end{align}
For the state and observables that give maximal violation in the case $p=0$, the violation is still visible as long as 
\begin{align}
    p \lesssim 0.24 \, . 
\end{align}

\subsection{Non-projective measurements}
Let us now suppose that instead of the projective measurements
\begin{align}
    A_1 & = S_x \,  , \qquad \qquad \qquad \quad \, \, \, \,  A_2 = S_x \cos \f + S_y \sin \f \, , \nonumber \\
    B_1 & = S_x \cos \f + S_y \sin \f \, , \quad B_2 = -S_x \, 
\end{align}
that lead to the maximal violation of the inequality, some POVMs with elements 
\begin{align}
    & \{E_1^{(1)}, E_1^{(0)}, E_1^{(-1)} \} \, , \quad \{E_2^{(1)},E_2 ^{(0)}, E_2^{(-1)} \} \, , \nonumber \\
    & \{F_1^{(1)}, F_1^{(0)}, F_1^{(-1)} \} \, , \quad \{F_2^{(1)},F_2 ^{(0)}, F_2^{(-1)} \} \,
\end{align}
are actually measured. Let us also suppose that instead of the theoretical commutators $[A_1, A_2]=: i  S_z \sin \f  =: i  A_3 \sin \f$ and $[B_1, B_2] =: i S_z \sin \f   =: i B_3 \sin \f$, some POVMs with elements $\{E_3^{(1)},E_3^{(0)}, E_3^{(-1)} \}$ and $\{F_3^{(1)}, F_3^{(0)}, F_3^{(-1)} \}$ are actually measured.  For this, we write the projective decomposition of the single-particle observables 
\begin{align}
    A_j = \sum_{a \in \{1,0,-1\}} a \, P_j^{(a)} = P_j^{(1)} - P_j^{(-1)} \, , \quad  B_j = \sum_{b \in \{1,0,-1\}} b \, Q_j^{(b)} = Q_j^{(1)} - Q_j^{(-1)} \, ,
\end{align}
where $P_j^{(a)}$ ($Q_j^{(b)}$) is the projector onto the subspace associated with eigenvalue $a$ ($b$) for the observable $A_j$ ($B_j$). Then, we let the POVM elements be $\e$-close to the corresponding projectors, that is,
\begin{align}
    E_j^{(a)} = P_j^{(a)} + \e C_j^{(a)} \, , \quad
    F_j^{(b)} = Q_j^{(b)} + \e D_j^{(b)}  \, ,
\end{align}
where $\e > 0$ and $C_j^{(a)}$ and $D_j^{(b)}$ have norm 1. Completeness of the POVMs implies
\begin{align}
    C_j^{(1)} + C_j^{(0)} + C_j^{(-1)} = 0 \, , \quad D_j^{(1)} + D_j^{(0)} + D_j^{(-1)} = 0 \, .
\end{align}
Hermiticity of the POVM elements implies hermiticity of the $C_j$'s and $D_j$'s. Finally we have positive semi-definiteness of the POVM elements:
\begin{align}
    E_j^{(a)} & = P_j^{(a)} + \e C_j^{(a)} \geq 0 \, , \nonumber \\
    F_j^{(b)} & = Q_j^{(a)} + \e D_j^{(b)} \geq 0 \, .
\end{align}
Then Eq. \eqref{eq:favgq} becomes
\begin{align}
    f = &    \avg{a_1^2} + \frac{2}{3} \avg{a_1 \ox a_1} - \frac{4}{3} \avg{a_1}^2     \nonumber  \\
    & +    \avg{b_1^2} + \frac{2}{3} \avg{b_1 \ox b_1}   - \frac{4}{3} \avg{b_1}^2\nonumber  \\
    & + \frac{2}{3} \Big( \avg{a_1 \ox b_1} - 2 \avg{a_1} \avg{b_1} \Big)      \nonumber  \\
    & +  \avg{a_2^2} + \frac{2}{3} \avg{a_2 \ox a_2} - \frac{4}{3} \avg{a_2}^2      \nonumber  \\
    & +   \avg{b_2^2}  + \frac{2}{3} \avg{b_2 \ox b_2}- \frac{4}{3} \avg{b_2}^2 \nonumber  \\
    & - \frac{2}{3} \Big( \avg{a_2 \ox b_2} - 2 \avg{a_2} \avg{b_2} \Big)     \nonumber  \\ 
    & -  \, \big| \sin \f \langle a_3 \rangle \big| -  \big| \sin \f \langle b_3 \rangle  \big|  \, .
\end{align}
We have
\begin{align}
    \avg{a_1} & = \sum_a a \, \tr \r E_1^{(a)}  \nonumber \\
    & = \tr \r (E_1^{(1)}-E_1^{(-1)} ) \nonumber \\
    & =  \tr \r (P_1^{(1)} + \e C_1^{(1)} - P_1^{(-1)} - \e C_1^{(-1)})  \nonumber \\
    & = \tr \r (A_1 + \e C_1^{(-)})   \nonumber \\
    & = \avg{A_1} + \e \avg{C_1^{(-)}} \, ,
\end{align}
and similarly
\begin{align}
     \avg{b_1} & =\avg{B_1} + \e \avg{D_1^{(-)}}  \, ,\nonumber \\
     \avg{a_2} & =\avg{A_2} + \e \avg{C_2^{(-)}} \, , \nonumber \\
     \avg{b_2} & =\avg{B_2} + \e \avg{D_2^{(-)}} \, , \nonumber \\
     \avg{a_3} & =\avg{A_3} + \e \avg{C_3^{(-)}} \, , \nonumber \\
     \avg{b_3} & =\avg{B_3} + \e \avg{D_3^{(-)}} \, .
\end{align}
On the other hand,
\begin{align}
    \avg{a_1^2} & = \sum_a a^2 \, \tr \r E_1^{(a)} \nonumber \\
    & = \tr \r ( E_1^{(1)} + E_1^{(-1)} )  \nonumber \\
    & = \tr \r ( P_1^{(1)} + \e C_1^{(1)} + P_1^{(-1)} + \e C_1^{(-1)} )\nonumber \\
    & = \tr \r A_1^2 + \e \tr \r C_1^{(+)} \nonumber \\
    & = \avg{A_1^2} + \e \avg{C_1^{(+)}}  \, ,
\end{align}
and similarly 
\begin{align}
    \avg{b_1^2} & = \avg{B_1^2} + \e \avg{D_1^{(+)}}  \, , \nonumber \\
    \avg{a_2^2} & = \avg{A_2^2} + \e \avg{C_2^{(+)}}  \, , \nonumber \\
    \avg{b_2^2} & = \avg{B_2^2} + \e \avg{D_2^{(+)}}  \, .
\end{align}
Finally,
\begin{align}
    \avg{a_1 \ox b_1} & =  \sum_{a,b} a b \, \tr \r E_1^{(a)} \ox F_1^{(b)}  \nonumber \\
    & = \tr \r ( E_1^{(1)} - E_1^{(-1)}) \ox (F_1^{(1)} - F_1^{(-1)}) \nonumber \\
    & = \tr \r ( P_1^{(1)} + \e C_1^{(1)} - P_1^{(-1)} - \e C_1^{(-1)}) \ox ( Q_1^{(1)} + \e D_1^{(1)} - Q_1^{(-1)} - \e D_1^{(-1)})  \nonumber \\
    & = \tr \r ( A_1 + \e C_1^{(-)} )\ox ( B_1 + \e D_1^{(-)} ) \nonumber \\
    & = \avg{A_1\ox B_1} +  \e \avg{A_1 \ox D_1^{(-)}} + \e \avg{C_1^{(-)} \ox B_1} + \e^2 \avg{C_1^{(-)} \ox D_1^{(-)}} \, ,
\end{align}
and similarly
\begin{align}
    \avg{a_1 \ox a_1} & =  \avg{A_1\ox A_1} + 2  \e \avg{A_1 \ox C_1^{(-)}}  + \e^2 \avg{C_1^{(-)} \ox C_1^{(-)}} \nonumber \\
    \avg{b_1 \ox b_1} & =  \avg{B_1\ox B_1} + 2  \e \avg{B_1 \ox D_1^{(-)}}  + \e^2 \avg{D_1^{(-)} \ox D_1^{(-)}} \nonumber \\
    \avg{a_2 \ox b_2} & =  \avg{A_2\ox B_2} +  \e \avg{A_2 \ox D_2^{(-)}} + \e \avg{C_2^{(-)} \ox B_2} + \e^2 \avg{C_2^{(-)} \ox D_2^{(-)}} \, , \nonumber \\
    \avg{a_2 \ox a_2} & =  \avg{A_2\ox A_2} + 2  \e \avg{A_2 \ox C_2^{(-)}}  + \e^2 \avg{C_2^{(-)} \ox C_2^{(-)}} \nonumber \\
    \avg{b_2 \ox b_2} & =  \avg{B_2\ox B_2} + 2  \e \avg{B_2 \ox D_2^{(-)}}  + \e^2 \avg{D_2^{(-)} \ox D_2^{(-)}} \, .    
\end{align}
For the particular case with the state $\r = \ket{\psi} \bra{\psi}$ with $\ket{\psi} =  c_{00} \ket{00}  + c_{02} \big( \ket{02} +  \ket{20} \big) + c_{11} \ket{11} + c_{22} \ket{22}$ and the observables $A_j = \cos \j_j^{(A)} S_x + \sin \j_j^{(A)} S_y$ and $B_j = \cos \j_j^{(B)} S_x + \sin \j_j^{(B)} S_y$ (with the numerical values given in \eqref{eq:cijnumerical} and \eqref{eq:anglesnumerical}) , we have 
\begin{align}
    \avg{a_1} & =   \e \avg{C_1^{(-)}} \, , \nonumber \\
     \avg{b_1} & = \e \avg{D_1^{(-)}}  \, ,\nonumber \\
     \avg{a_2} & = \e \avg{C_2^{(-)}} \, , \nonumber \\
     \avg{b_2} & = \e \avg{D_2^{(-)}} \, , \nonumber \\
     \avg{a_3} & =|c_{00}|^2 -|c_{22}|^2 + \e \avg{C_3^{(-)}} \, , \nonumber \\
     \avg{b_3} & =|c_{00}|^2 -|c_{22}|^2 + \e \avg{D_3^{(-)}} \, ,
\end{align}
\begin{align}
    \avg{a_1^2} & = \frac{1 + c_{11}^2 + 2 c_{02} \re (c_{00} + c_{22})}{2} + \e \avg{C_1^{(+)}}  \, , \nonumber \\
    \avg{b_1^2} & = \frac{1 + c_{11}^2 + 2 c_{02} \re (e^{2 i \f} c_{00} + e^{-2 i \f} c_{22})}{2} + \e \avg{D_1^{(+)}}  \, , \nonumber \\
    \avg{a_2^2} & = \frac{1 + c_{11}^2 + 2 c_{02} \re (e^{2 i \f} c_{00} + e^{-2 i \f} c_{22})}{2} + \e \avg{C_2^{(+)}}  \, , \nonumber \\
    \avg{b_2^2} & = \frac{1 + c_{11}^2 + 2 c_{02} \re (c_{00} + c_{22})}{2} + \e \avg{D_2^{(+)}}  \, .
\end{align}
and 
\begin{align}
    \avg{a_1 \ox b_1} & = c_{11} \left[ 2 c_{02} \cos \f + \re (e^{i \f} c_{00} + e^{-i \f } c_{22} )\right] +  \e \avg{A_1 \ox D_1^{(-)}} + \e \avg{C_1^{(-)} \ox B_1} + \e^2 \avg{C_1^{(-)} \ox D_1^{(-)}} \, , \nonumber \\
    \avg{a_1 \ox a_1} & =  c_{11} \left[ 2c_{02} + \re ( c_{00} + c_{22}) \right] + 2  \e \avg{A_1 \ox C_1^{(-)}}  + \e^2 \avg{C_1^{(-)} \ox C_1^{(-)}} \, , \nonumber \\
    \avg{b_1 \ox b_1} & = c_{11} \left[ 2 c_{02} + \re (e^{2 i \f} c_{00} + e^{-2 i \f} c_{22} ) \right] + 2  \e \avg{B_1 \ox D_1^{(-)}}  + \e^2 \avg{D_1^{(-)} \ox D_1^{(-)}} \, , \nonumber \\
    \avg{a_2 \ox b_2} & = - c_{11} \left[ 2 c_{02} \cos \f + \re (e^{i \f} c_{00} + e^{-i \f } c_{22} )\right] +  \e \avg{A_2 \ox D_2^{(-)}} + \e \avg{C_2^{(-)} \ox B_2} + \e^2 \avg{C_2^{(-)} \ox D_2^{(-)}} \, , \nonumber \\
    \avg{a_2 \ox a_2} & =  c_{11} \left[ 2 c_{02} + \re (e^{2 i \f} c_{00} + e^{-2 i \f} c_{22} ) \right] + 2  \e \avg{A_2 \ox C_2^{(-)}}  + \e^2 \avg{C_2^{(-)} \ox C_2^{(-)}} \, , \nonumber \\
    \avg{b_2 \ox b_2} & =  c_{11} \left[ 2c_{02} + \re ( c_{00} + c_{22}) \right] + 2  \e \avg{B_2 \ox D_2^{(-)}}  + \e^2 \avg{D_2^{(-)} \ox D_2^{(-)}} \, . 
\end{align}
Optimizing for the worst case scenario with the $C_j^{(a)}$'s and $D_j^{(a)}$'s Hermitian norm-1 operators, we find that the violation of the inequality is still visible as long as
\begin{align}
    \e \lesssim 0.02 \, .
\end{align}

\end{document}